\documentclass{article}

\usepackage[english]{babel}

\usepackage[letterpaper,top=2cm,bottom=2cm,left=2.5cm,right=2.5cm,marginparwidth=1.75cm]{geometry}

\usepackage{amsmath}
\usepackage{graphicx}
\usepackage[colorlinks=true, allcolors=blue]{hyperref}

\usepackage{amssymb}
\usepackage{url}
\usepackage{mathtools}

\usepackage{xcolor}
\newcommand{\kamcomment}[1]{\textcolor{blue}{#1}}

\usepackage{amsmath}
\usepackage{bm}
\usepackage{caption}
\usepackage{subcaption}
\usepackage{booktabs}
\usepackage{algorithm}
\usepackage{algpseudocodex}
\usepackage{amsthm}
\theoremstyle{remark}
\newtheorem*{remark}{Remark}

\title{A Regularized Ensemble Kalman Filter for Stochastic Phase Field Models of Brittle Fracture}

\author{Lucas Hermann$^1$ \and Ralf Jänicke$^2$ \and Knut Andreas Meyer$^3$ \and Ulrich Römer$^1$}
\date{%
    \small$^1$TU Braunschweig, Institute for Acoustics and Dynamics\\%
    \small$^2$TU Braunschweig, Institute for Applied Mechanics\\
    \small$^3$Chalmers University of Technology, Dept. of Mechanical Engineering\\[2ex]%
    09.03.2026
}

\begin{document}
\maketitle



\begin{abstract}
The phase-field approach to brittle fracture provides a continuum framework for modeling crack initiation and propagation without explicit representation of discrete crack surfaces, provided the spatial discretization is fine enough to resolve the regularization length scale.
%
However, uncertain local material parameters due to material defects can strongly influence simulation results, such as crack paths and remaining structural strength.
At the same time, the ability to continuously monitor structures using sensors allows complementing modeling predictions with, e.g., displacement measurements. 

In this contribution, we connect these two complementary sources of information and present a Bayesian inference procedure that allows updating the current model state with incoming sensor data. We construct a Bayesian prior for the model state (both displacements and phase-field) and employ an ensemble Kalman filter (EnKF) to perform the update. In the EnKF, the update is computed by performing a Kalman shift on each ensemble member. Since the standard EnKF may produce assimilated states that violate common modeling assumptions, we present a phase field-based regularization technique as a proximal step correction toward model-consistent updates. 1D and 2D numerical examples demonstrate the performance and accuracy of the proposed method and show that the updated state matches the ground truth reasonably well.  
Unlike traditional Bayesian inversion techniques, which have already been applied to brittle fracture, we infer not the model parameters but the model state, i.e., the displacement field and the phase-field. Although only displacements are observed, the strong correlation between both fields also allows inference of the posterior phase-field.
\end{abstract}



\section{Introduction}
\label{sec1}

Phase field fracture modeling was originally proposed to simulate crack propagation in brittle materials \cite{francfortRevisitingBrittleFracture1998, mieheThermodynamicallyConsistentPhasefield2010}, and has later been extended to ductile fracture \cite{ambatiPhasefieldModelingDuctile2015} and fatigue crack growth \cite{kalinaOverviewPhasefieldModels2023}. The gradient-regularization ensures mesh-independent results in the case of a sufficiently fine spatial discretization. However, crack propagation is highly dependent on variations in boundary conditions and local material defects, both in physical experiments and models. 
Given that material parameters and initial conditions are often only partially known, Uncertainty Quantification (UQ) methods are employed to establish a probabilistic forward problem \cite{gerasimovStochasticPhasefieldModeling2020,nagarajaDeterministicStochasticPhasefield2023}.

In this context, the forward problem entails propagating uncertainty in the input parameters to model outputs. Input parameters can be related to material properties, initial conditions, and boundary conditions.
The goal is then to determine the induced probability distribution for model outputs, such as displacement and the phase field and the corresponding crack paths. In this context, multivariate Gaussian approximations for the output densities are appealing to reduce the computational cost. These Gaussian distributions are defined via a mean and variance for each point of the domain, as well as two-point correlation functions and the correlations across fields. 

For non-linear phase field problems, the uncertainty in the output can be substantial. To mitigate this uncertainty, real-world data can be collected, for instance, through Digital Image Correlation (DIC) \cite{chu1985applications}. In a Bayesian context, prior distributions can be conditioned on this data, see, e.g. \cite{girolamiStatisticalFiniteElement2021}, effectively assimilating the data into the model. For a specific structure of the data model, this results in a new Gaussian posterior that has a reduced variance, particularly around sensor locations. 

The phase field fracture model operates using a quasi-static time-stepping regime, incrementally increasing the load to trace the evolving damage. In some cases, measurement data may be available at various time steps or loads during the simulation, requiring conditioning of the time-evolving probability distribution on incoming data. This process can be managed through Kalman filtering, where the model is advanced until the time step with available data. The solution is then conditioned on the measured data, and the resulting posterior distribution serves as the initial condition for subsequent uncertainty propagation. For non-linear models, the Extended Kalman Filter (EKF) is typically employed to linearize the problem. Alternatively, the Ensemble Kalman Filter (EnKF) can be used for non-linear problems, operating in a Monte Carlo fashion, which reduces uncertainty over time as different members of the ensemble converge toward a common mean. The reader is referred to \cite{duffinStatisticalFiniteElements2021} and the references therein for a discussion.

Applying Kalman filtering to the phase field brittle fracture model presents several challenges. Since the phase field variable attempts to model a smeared but localized crack, fine discretization and the positioning of sensors around the crack path are needed. Additionally, the phase field variable is history-dependent, requiring measures to prevent unphysical solutions during the conditioning process. The aforementioned non-linearities further complicate the filtering approach. 

This paper aims to provide a data-corrected solution to the phase field fracture problem, particularly when dealing with non-unique crack path solutions. These may occur if the phase field is highly sensitive to mesh parameters, initial conditions or material configurations \cite{gerasimovStochasticPhasefieldModeling2020}. Our goal is to infer the phase field solution based on data for the displacement, leveraging the model-based correlation between both fields. 

It is important to note that there exists other works on data assimilation procedures, applied to numerical fracture problems. For instance, \cite{wuParameterIdentificationPhasefield2021} derived a Bayesian approach to update scalar parameters of the phase field model, albeit not the complete high-dimensional state as targeted in this publication. The authors in \cite{pulikkathodiRealtimeInverseCrack2023} employ a linear elastic fracture mechanics approach and use it in conjunction with an EKF. The method parametrizes the crack, hence e.g. the length or the angle of the crack surface take a certain value, given a load increment. These parameters are subject to uncertainty and updated throughout time with the EKF. The authors state that, so far, only cracks in form of a straight line can be parametrized and updated in the given framework. 
In \cite{rochinhaMonitoringHydraulicFractures2010}, the authors make use of an hydraulic fracture model, the parameters of which are updated using, again, an EKF. Here, the corresponding crack surface, i.e. the shape of the crack path, is parametrized with few parameters which are updated with the EKF. 
The authors of \cite{kunchamOnlineModelbasedFatigue2022} focus on fatigue fracture. The model is based on XFEM and three crack parameters, stemming from Paris' law for fatigue crack growth, are updated with an EKF.
In \cite{pingHistoryMatchingFracture2013}, the authors use an EnKF to update fracture length and angle parameters of a 2D fracture field featuring multiple straight cracks.
The mentioned references have in common that they use a Kalman Filter and parametric fracture models. The downside of these parametric fracture models, albeit being rather computationally cheap, is a lack in flexibility concerning the crack path compared to phase field models. In many cases, the overall shape and rough path of the expected crack need to be known beforehand. Geometrically complex crack surfaces are very hard to model. For XFEM, specifically, tracking a crack in three dimensions is challenging. The phase field approach to brittle fracture, as adopted in this paper, however, doesn't assume a certain geometry of the crack. An extension to three dimensions is rather simple. Instead of updating only few parameters, the complete state of the phase field and displacement is updated using an EnKF. This makes it possible to gather real world data from complex crack paths and, in theory, end up with a model solution much closer to reality compared to the parametric models. The differences of phase field fracture models to a parametric one (in that case XFEM) are explained in great detail in \cite{cerveraComparativeReviewXFEM2022}.

The paper is structured as follows: The next chapter introduces phase field brittle fracture modeling, with a focus on the micromorphic phase field model \cite{bharaliMicromorphicPhasefieldModel2024}. Following this, the stochastic forward problem is presented, resulting in the prior state suitable for data assimilation. The subsequent section addresses the stochastic data model as well as the filtering. We put a special focus on the regularization of the posterior which is necessary in order to enforce physically relevant properties. A 1D numerical example is developed alongside the theory to illustrate the procedure. In the last chapter, a 2D example is used to demonstrate the ideas in a more relevant setting.


\subsection{Notation}
Before proceeding to the phase field model and Kalman filtering techniques, we first comment on the notation used throughout this study. Scalars (scalar fields) are denoted by Roman type, while vectors (vector fields) are represented with bold-italic characters. Additionally, blackboard bold characters are reserved for function spaces and sets (e.g. $\mathbb{R}$). After discretization, bold-italic characters are used again for vectors and matrices. 
A Gaussian random vector is denoted as $\bm{a} \sim \mathcal{N}(\bar{\bm{a}},\bm{C}_{\bf{a}})$, with mean vector $\bar{\bm{a}}$ and covariance matrix $\bm{C}_{\bm{a}}$. We use $p(\bm{a})$ to denote the (multivariate) probability density function of $\bm{a}$ and $p(\bm{a}|\bm{b})$ to denote the density, conditional on $\bm{b}$. We use the same symbol for a random variable/vector and the respective realization, since the meaning is clear from the context. The quantity $\|\bm{a}\|_{\bm{C}}^2 = \bm{a}^\top \bm{C}^{-1} \bm{a}$ denotes a weighted Euclidean norm. We will employ indices as follows: $\bm{a}_{n,k}$ denotes the $k$-th iterate in the Newton-Raphson iteration at displacement increment $n$ and $\bm{a}_{n}$ refers to the final iterate. During Kalman filtering we distinguish between the forecasted state $\bm{a}^{\text{F}}$ and the assimilated state $\bm{a}^{\text{A}}$ from data. In the context of the ensemble Kalman filter, $\bm{a}^{\text{F}/\text{A}}_{n,i}$ denote the forecasted/assimilated state at increment $n$ of the $i$-th ensemble member. Finally, we distinguish between a function and the value it takes, e.g., $\hat{E}$ refers to a function, whereas $E$ is the value it takes given specific arguments.

\section{Phase field brittle fracture}
\label{sec2}

\begin{figure}[ht!]

    \begin{minipage}[b]{.48\linewidth}
    \centering
    \includegraphics[page=3]{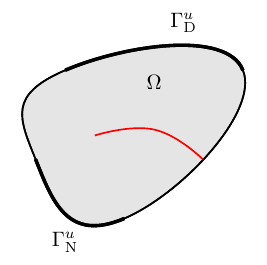}
    \subcaption{a) A crack is represented by a sharp fracture surface, which is challenging to handle numerically.}
        \label{fig:Fig1a}

  \end{minipage}\hfill
  \begin{minipage}[b]{.48\linewidth}
    \centering
 \includegraphics[page=4]{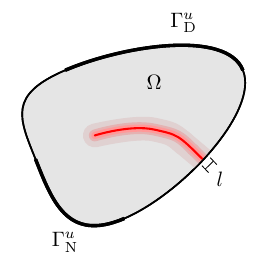}
   \subcaption{b) Through phase field regularization, the fracture surface is effectively approximated by a smooth field. }
        \label{fig:Fig1b}
  \end{minipage}
  \caption{Phase field regularization of a fracture in a continuum, based on \cite{bharaliMicromorphicPhasefieldModel2024}}
\end{figure}

The phase field modeling approach summarized in this chapter has been put forth in \cite{bharaliMicromorphicPhasefieldModel2024,delorenzisNumericalImplementationPhaseField2020,khiminModelingDiscretizationOptimization2023}. Let $\Omega \subset \mathbb{R}^{d}$, $d=1,2$, be an open and bounded domain, occupied by a linear elastic continuum.  The boundary is separated into a Dirichlet boundary $\Gamma_{\text{D}}^u$ and, besides the natural boundary conditions, a prescribed Neumann boundary $\Gamma_{\text{N}}^u$, such that $\Gamma_{\text{D}}^u \cap \Gamma_{\text{N}}^u = \varnothing$. For the phase field itself, no boundary conditions are prescribed, hence only homogeneous Neumann conditions are involved. Within the continuum, cracks can develop, as visible in Figure \ref{fig:Fig1a}.

A sharp crack results in a discontinuity within the continuum along the time-evolving crack surface $\Gamma_{\text{c}}$. The phase field method for brittle fracture aims to regularize the discontinuity in terms of the phase field variable $\varphi (\bm{x},t) \in [0,1]$, where $0$ signifies an undamaged material and $1$ a fully fractured one. The length scale parameter $\ell \in \mathbb{R}^+$ controls the width of the resulting regularized crack, as visible in Figure \ref{fig:Fig1b}.
The phase field method for brittle fracture is based on minimising an energy functional 
\begin{equation}
\label{eq:ppEnFunc}
\begin{split}
    E(\bm{u}, \varphi) = \int_{\Omega} g(\varphi)\Psi^+ (\hat{\bm{\epsilon}}(\bm{u}))\; \text{d}\Omega + \int_{\Omega} \Psi^- (\hat{\bm{\epsilon}}(\bm{u}))\; \text{d}\Omega - \int_{\Gamma_{\text{N}}^u} \bm{t}_{\text{N}} \cdot \bm{u} \;\text{d}\Gamma & \\
    + \int_{\Omega} \frac{G_{\text{c}}}{c_{\text{w}} \ell} \left( w(\varphi) + \ell^2 |\nabla \varphi|^2 \right) \text{d}\Omega &,
    \end{split}
\end{equation}
in which $\Psi^{\pm} \left(\hat{\bm{\epsilon}}(\bm{u})\right)$ is the split elastic strain energy density, where $\Psi^{+}$ is the fracture driving part. Moreover, $\hat{\bm{\epsilon}}(\bm{u})$ is the symmetric part of the displacement gradient and $\bm{u}(\bm{x},t)$ the displacement.
In the numerical examples, the volumetric-deviatoric split is used and we have for $\Psi^{\pm}$ 
\begin{equation}
  \Psi^{+} = \frac{1}{2} K \langle\text{tr}(\bm{\epsilon})\rangle^2 + \mu  (\bm{\epsilon}_\text{dev} : \bm{\epsilon}_\text{dev}), \quad \Psi^{-} = \frac{1}{2} K \langle-\text{tr}(\bm{\epsilon})\rangle^2,
\end{equation}
The boundary term over the Neumann boundary $\Gamma_{\text{N}}^u$ models an applied traction. 
Finally, we use the so-called AT2 fracture model with
\begin{equation}
\label{eqn:degradation}
    g(\varphi) = (1-\varphi)^2, \quad c_w = 2, \quad w(\varphi) = \varphi^2
\end{equation}
where $g$ is the degradation function, $w$ the local dissipation function, and $c_w$ the associated scaling to make $G_{\text{c}}$ the fracture energy.
Notice that this is a quasi-static problem, hence the numerical problem will be treated as being stationary. Therefore, we omit time indices at this point of the derivation and will introduce these later when they become relevant for the EnKF. For the moment it is sufficient to introduce $\varphi'$ as the phase field of the previous time/loading step.

In this paper, a modified version of the phase field problem will be used, namely the micromorphic approach according to \cite{bharaliMicromorphicPhasefieldModel2024}. The energy functional parametrization is then extended by the micromorphic variable, $d$, and a term penalizing the difference between the phase field $\varphi$ and $d$ is added, resulting in 
\begin{equation}
\begin{split}
    E = \hat{E}(\bm{u}, \varphi, d) = \int_{\Omega} g(\varphi)\Psi^+ (\hat{\bm{\epsilon}}(\bm{u}))\; \text{d}\Omega + \int_{\Omega} \Psi^- (\hat{\bm{\epsilon}}(\bm{u}))\; \text{d}\Omega - \int_{\Gamma_{\text{N}}^u} \bm{t}_{\text{N}} \cdot \bm{u} \;\text{d}\Gamma & \\
    + \int_{\Omega} \frac{G_{\text{c}}}{c_{\text{w}} \ell} \left( w(\varphi) + \ell^2 |\nabla d|^2 \right) \text{d}\Omega + \int_{\Omega} \frac{\alpha}{2}(\varphi - d)^2 \text{d}\Omega &.
\end{split}
\label{eq:energy_micromorphic}
\end{equation}
By introducing the function spaces 
\begin{subequations}
    \begin{align}
    \mathbb{U} &= \{\bm{u} \in (H^1(\Omega))^{2}\; | \; \bm{u} = \bm{u}_{\text{D}} \quad \text{on} \; \Gamma_{\text{D}}^u\},\\
    \mathbb{U}_0 &= \{\bm{u} \in (H^1(\Omega))^{2}\; | \; \bm{u} = \bm{0} \quad \text{on} \;  \Gamma_{\text{D}}^u\},\\
    \mathbb{D} &= H^1(\Omega),
    \end{align}
\end{subequations}
the Euler-Lagrange equations, related to \eqref{eq:energy_micromorphic}, read: find $(\bm{u},d,\varphi) \in \mathbb{U} \times \mathbb{D} \times L^2(\Omega)$ such that
\begin{align}
    \int_{\Omega} g(\varphi) \bm{\sigma}^+ \colon \hat{\bm{\epsilon}}(\delta\bm{u})\; \text{d}\Omega + \int_{\Omega} \bm{\sigma}^- \colon \hat{\bm{\epsilon}}(\delta\bm{u}) \;\text{d}\Omega -  \int_{\Gamma_{\text{N}}^u} \bm{t}_{\text{N}} \cdot \delta\bm{u} \;\text{d}\Gamma &= 0    \qquad \forall \delta\bm{u} \in \mathbb{U}_0, \label{eqn:strong1}\\
    \int_{\Omega} \frac{2G_{\text{c}}\ell}{c_{\text{w}}} \nabla d \cdot \nabla \delta d \; \text{d}\Omega -  \int_{\Omega} \alpha (\varphi - d) \delta d \;\text{d}\Omega &= 0    \qquad \forall \delta d \in \mathbb{D},\label{eqn:strong2}\\
    \int_{\Omega} \delta\varphi \left(g'(\varphi)\Psi^+ (\hat{\bm{\epsilon}}(\bm{u})) + \frac{G_{\text{c}}}{c_{\text{w}} \ell} w'(\varphi)) + \alpha(\varphi - d)\right) \text{d}\Omega &= 0    \qquad \forall \delta \varphi \in L^2(\Omega)\;. \label{eqn:weak3}
\end{align}
Equation \eqref{eqn:weak3} implies that 
\begin{align}
    g'(\varphi)\Psi^+ (\hat{\bm{\epsilon}}(\bm{u})) + \frac{G_{\text{c}}}{c_{\text{w}} l} w'(\varphi) + \alpha(\varphi - d) = 0, \label{eqn:stron3local}
\end{align}
almost everywhere\footnote{That is except on sets of Lebesgue measure zero.} in $\Omega$, however, we will assume additional regularity so that \eqref{eqn:stron3local} holds pointwise in $\Omega$. 
The constraints on $\varphi$, irreversibility, $\dot{\varphi}\geq 0$, and bounds, $0 \leq \varphi \leq 1$, can then be enforced locally. With the chosen AT2 fracture model, Equation \eqref{eqn:stron3local} can then be solved analytically as
\begin{equation}
\label{eqn:phiFromd}
    \varphi = \hat{\varphi}(\bm{u},d,\varphi',\ell) = \text{min} \left(\text{max}  \left(   \frac{2\Psi^+(\hat{\bm{\epsilon}}(\bm{u})) + \alpha d }{2\Psi^+(\hat{\bm{\epsilon}}(\bm{u})) + \alpha + \frac{G_\text{c}}{\ell}}, \varphi' \right) , 1   \right),
\end{equation}
where the dependence on quantities $(\bm{u},d,\varphi',\ell)$ is highlighted, since these will be modified later on as part of the Kalman filter. 
\subsection*{FEM for PF fracture}
The phase-field fracture problem is solved with the Finite Element Method (FEM), introducing the approximations
\begin{align}
    \bm{u}(\bm{x}) \approx \bm{u}_h(\bm{x}) =  \sum_i \bm{\phi}^u_i(\bm{x}) a^u_i, \quad 
    d(\bm{x}) \approx d_h(\bm{x}) = \sum_i \phi^d_i(\bm{x}) a^d_i
\end{align}
such that the energy, $E \approx \hat{E}(\bm{u}_h, d_h;\varphi_h',\ell) = \hat{E}_{h}(\bm{a}^u, \bm{a}^d;\bm{\varphi}',\ell)$. Note that because of $\varphi_h = \hat{\varphi}(\bm{u}_h,d_h,\varphi_h';\ell)$ we have rewritten $\hat{E}(\bm{u}_h, \varphi_h, d_h) = \hat{E}(\bm{u}_h, d_h;\varphi_h',\ell)$, where we now explicitly state the dependence on $\varphi_h',\ell$. This notation will be required to introduce the Kalman filter regularization. 
Minimizing $\hat{E}_h$ leads to 
\begin{align}
    R^u_i(\bm{a}^u, \bm{a}^d; \bm{\varphi}', \ell) &= \int_{\Omega} \hat{\bm{\epsilon}}(\bm{\phi}^u_i) : \left[g(\varphi_h) \bm{\sigma}^+ + \bm{\sigma}^-\right]\; \text{d}\Omega -  \int_{\Gamma_{\text{N}}^u} \bm{\phi}^u_i \cdot \bm{t}_{\text{N}} \;\text{d}\Gamma = 0, \label{eq:discrete_residual_1}\\
    R^d_i(\bm{a}^u, \bm{a}^d; \bm{\varphi}', \ell) &= \int_{\Omega} \nabla \phi^d_i \cdot \left[\frac{2G_{\text{c}}\ell}{c_{\text{w}}} \nabla d_h \right]  \; \text{d}\Omega -  \int_{\Omega} \phi^d_i \alpha (\varphi_h - d_h)\;\text{d}\Omega = 0 \label{eq:discrete_residual_2}.
\end{align}
Following \cite{bharaliMicromorphicPhasefieldModel2024}, we adopt convexification via extrapolation \cite{heisterPrimaldualActiveSet2015}, using the extrapolated micromorphic field,
\begin{equation}
    \check{\bm{a}}^d = \bm{a}^d_{n-1} + \Delta_{t_n}\frac{(\bm{a}^d_{n-1} - \bm{a}^d_{n-2})}{\Delta_{t_{n-1}}}, \quad \check{d}_h = \sum_i \phi_i^d \check{a}_i^d,
\end{equation}
and $\varphi \approx \check{\varphi} = \hat{\varphi}_h(\bm{u}_h, \check{d}_h;\varphi_{n-1},\ell)$ in \eqref{eq:discrete_residual_1} and \eqref{eq:discrete_residual_2}. Note that we only indicate past indices ${n-1},n-2$ and not the current one $n$, to keep the notation lighter. We also use $\varphi$ instead of $\check{\varphi}$ in the following, keeping in mind that always $\check{d}_h$ is used here as an argument. 
With these definitions, we now collect the state variables (per quadrature point), $\bm{\varphi}$, unknowns, $\bm{a} = (\bm{a}^u, \bm{a}^d)$, and residuals, $\bm{R} = (\bm{R}^u, \bm{R}^d)$, such that we have the complete equation system
\begin{align}
    \bm{R}^u(\bm{a}^u, \check{\bm{a}}^d; \bm{\varphi}_{n-1}, \ell) &= 0, \\
    \bm{R}^d(\bm{a}^u, \bm{a}^d; \bm{\varphi}_{n-1}, \ell) &= 0, \label{eq:discrete_residual_n}
\end{align}
that we solve monolithically in each time step using the standard Newton-Raphson method. Note that \eqref{eq:discrete_residual_n} is linear in $\bm{a}^d$, as $\check{\bm{a}}^d$ is used to calculate $\varphi_h$.

\begin{remark}
    Solving the FE system of equations which is defined in this section is computationally very demanding. In \cite{kristensenPhaseFieldMethods2022}, the CPU hours were compared for different solution schemes. As we not only aim to solve a deterministic phase field problem, but a stochastic one based on an ensemble strategy, which potentially involves hundreds of parallel model evaluations, some precautions have to be taken to keep the CPU hours per ensemble member low and hence the stochastic evaluation feasible. 

    Firstly, a coarser mesh is applied in the regions where a crack is not expected in order to reduce the number of degrees of freedom of the resulting FE system of equations. Secondly, the monolithic solution strategy has been chosen deliberately over a staggered one, because it requires significantly less CPU hours \cite{kristensenPhaseFieldMethods2022}. Lastly, the Volumetric-Deviatoric energy split 
    is chosen
    over the spectral decomposition, as that split is computationally more demanding than the Volumetric-Deviatoric split. 
\end{remark}

\subsection{Deterministic solution of the phase field problem/ True underlying solution}
\label{DetPP}
\begin{figure}[ht!]

    \begin{minipage}[b]{.55\linewidth}
    \centering
    \includegraphics[page=5]{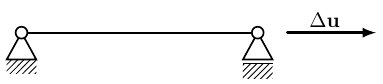}
    \captionof{figure}{1D numerical experiment}
        \label{fig:Figpp1d}

  \end{minipage}\hfill
  \begin{minipage}[b]{.45\linewidth}
    \centering
\begin{tabular}{@{}ll@{}}
\toprule
\textbf{Parameter} & \textbf{Value}      \\ \midrule
Fracture Model & AT2                 \\
$E$            & $210000$ {[}N/$\text{mm}^2${]} \\
$\nu$          & $0.3$ {[}-{]}       \\
$G_\text{c}$          & $2.7$ {[}N/mm{]}    \\
$l$            & $2.5e-2$ {[}mm{]}   \\
$\alpha$       & $\beta G_\text{c} /\ell$      \\ \bottomrule
\end{tabular}
    \captionof{table}{Model parameters for the 1D tension rod example}
    \label{tab:1dpar}
  \end{minipage}
\end{figure}

Solving the phase field problem for a fixed set of parameters yields a deterministic solution. This solution can in the further process serve as a reference solution from which data can be observed.
We want to explain the concepts developed in this paper along with a simple example. To this end, we introduce a solution to the phase field problem on a 1D domain. Figure \ref{fig:Figpp1d} illustrates the mechanical setup: A tension rod on $\Omega = [-1,1]$ in [mm] is pulled in a displacement controlled manner until failure.

\begin{figure}[ht!]
    \centering
    \includegraphics[]{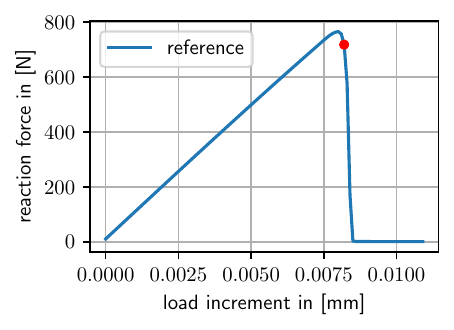}
    \caption{Reaction forces over pseudo time for the 1D numerical example. A linear elastic interval is followed by an abrupt crack and loss of remaining stiffness. The red dot marks the first load increment and reaction force at which data assimilation will take place in the later section.}
    \label{fig:react1d}
\end{figure}

With $N=200$ degrees of freedom for each $u$ and $d$, i.e. a mesh size of $h=1e-2\, \text{mm}$, the length scale is fixed to $l = 2.5e-2$. We use piecewise linear approximations for $u$ and $d$. While the left boundary is fixed, a Dirichlet boundary condition on the right boundary is used to prescribe a displacement increment of $\Delta u = 1e-4$ [mm] per pseudo time step. As the problem is 1D, the energy split does not apply here. Table \ref{tab:1dpar} lists all chosen parameters for the model. An initial value of $\varphi = 0.7$ for the phase field at a certain position, in this case $x_0 = 0.57$, i.e. an initial damage, serves as a nucleus for the development of a crack.
In Figure \ref{fig:1dprior}, the deterministic solution for both phase field and displacements is visible in blue for a single time step.
The whole time series can be condensed to a force displacement diagram as visible in Figure \ref{fig:react1d}. The time step marked in red is the one at which the displacement and phase field is shown in the plots of the next sections.


\section{Data Assimilation}
If both a predictive model and measured data exist, the process of connecting both, to ideally obtain improved predictions, is called data assimilation \cite{evensenDataAssimilationEnsemble2009}. 
Assimilating a normally distributed output from a linear model with measurement data that also follows a normal distribution can be achieved in closed-form via conditioning of multivariate Gaussian distribution, see \cite{rasmussenGaussianProcessesMachine2006}. In the time-dependent case, a Kalman filter can be used. Details on the most important flavours of the Kalman Filter and related techniques can be found in \cite{sanz-alonsoInverseProblemsData2023a}. The Kalman filter is very similar to the usual conditioning of a multivariate Gaussian, with the difference that the updated distribution serves as a prior in a subsequent time step. 
The standard Kalman filter is derived for linear problems. If the problem is non-linear, linearisation leads to the extended Kalman filter (EKF). Another method to treat the non-linearity is to use derivative-free sampling methods. Examples for sampling-based data assimilation methods are the Ensemble Kalman Filter (EnKF) \cite{evensenEnsembleKalmanFilter2003} and the Particle Filter \cite{gordonNovelApproachNonlinear1993}. In any case, a prior is needed and is obtained here by propagating uncertainties through the predictive model. 
The following introduction and discussion is based on \cite{duffinStatisticalFiniteElements2021, sanz-alonsoInverseProblemsData2023a}.

\subsection{Stochastic forward problem}
Many aspects of numerical modeling require making assumptions, handling a lack of knowledge and quantifying errors, originating e.g. from discretisation. In these cases, instead of using deterministic quantities, accounting and propagating uncertainties is crucial. 
For the given phase field problem, there are many sources of uncertainty. 
A prominent example is parametric uncertainty, e.g. in the boundary, initial conditions and material parameters which can be either scalar or fields. Another source of uncertainty is in the chosen initial damage field. 
In this paper, we will focus on the latter and introduce a random initial damage field.
Uncertainty in these input parameters will lead to uncertainty in the solution fields, as well. For a linear model, Gaussian distributions in the random coefficients will lead to a Gaussian solution. The EnKF, which is the data assimilation technique used in this work, propagates a particle ensemble through the model, which can be used to approximate the output statistics and obtain a prior. Hence, uncertainty propagation is carried out in a Monte Carlo (MC)-like fashion, where the phase field model is simulated repeatedly with varying realizations of the model inputs. 

\begin{figure}[ht!]
    \centering
    \includegraphics[width=12cm]{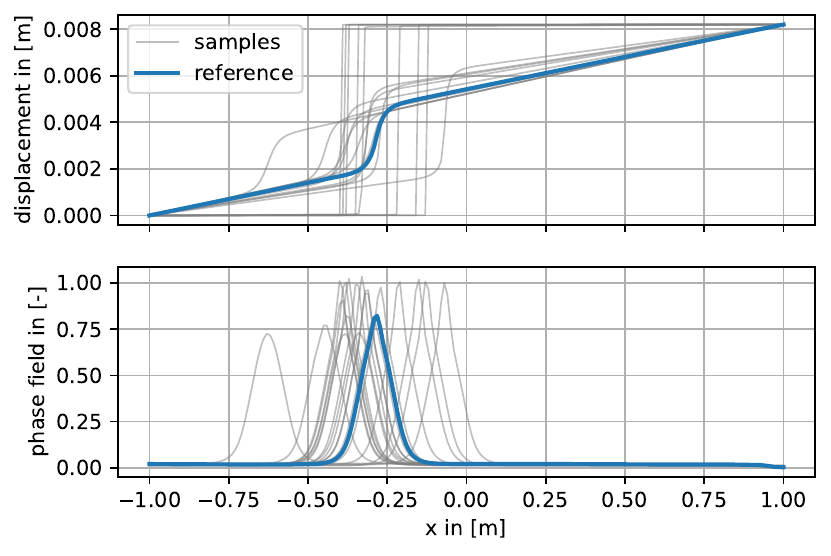}
    \caption{Solution to the 1D phase field brittle fracture problem at a single time step. The deterministic reference solution is visible in blue, whereas the prior ensemble is visible in grey. Some ensemble members indicate a complete fracture (phase field of $1$, displacement is a step function) while others only have a phase field smaller than $1$, hence there is some stiffness degradation but no complete fracture. The prior ensemble is the result of row $6$ in Algorithm \ref{alg:kalman}.}
    \label{fig:1dprior}
\end{figure}

In Figure \ref{fig:1dprior}, multiple samples of the stochastic phase field problem are shown. These arise from treating the initial value for the phase field as a random variable: The position and magnitude of the damage nucleus is different for each realisation, hence also the position and behaviour in time of the resulting crack.
The position is sampled from a normal distribution $x_0 \sim \mathcal{N}\left(-0.25, 0.12\right)$, whereas the initial magnitude of the phase field is sampled from a uniform distribution $\varphi_{\text{mag}} \sim \mathcal{U}(0.73,0.76)$. The shape of the initial damage is defined as a Gaussian with magnitude $\varphi_{\text{mag}}$, standard deviation $\sigma = 0.05$ and mean $x_0$.
From the ensemble, a Gaussian approximation can be computed, i.e. mean and covariance are calculated from the samples.
The next step towards data assimilation is to connect the prior ensemble with data in form of a data model.

\subsection{Data model}
The basis for assimilating model predictions with data is a model of the observations. A data model connects the phase field FE solution to the sensor data and the derivation in this section mainly follows \cite{girolamiStatisticalFiniteElement2021}.
First, based on \cite{kennedyBayesianCalibrationComputer2001}, we model the measurement data generating process as 
\begin{equation}
\label{eqn:datamodel}
    \bm{y}_n = \rho \bm{H} \bm{a}_n +\bm{\delta}_n +\bm{\eta}_n
\end{equation}
now with an explicit iteration index $n$, the measured data $\bm{y}_n \in \mathbb{R}^N$ at time $n \Delta_t$ and the linear observation map $\bm{H} \colon \mathbb{R}^M \to \mathbb{R}^N$  which maps from the high-dimensional state-space of the solution to the lower dimensional measurement space. The matrix can be constructed by evaluating the FE basis functions at the sensor locations. In general, $\bm{H}$ is constructed by discretising the respective operator relating the FE solution to measurable quantities. Moreover, $\bm{\eta}_n \sim \mathcal{N}(\bm{0}, \sigma^2_n \bm{I})$ models measurement noise and 
    \begin{equation*}
   \bm{a}_{n} \sim \mathcal{N}\left( \bar{\bm{a}}_{n},\bm{C}_{a;n} \right)
 \end{equation*} is the FE solution vector containing both displacements $\bm{u}$ and the micromorphic field $\bm{d}$. At every time/load step, data can be generated. Note that measurements will only be taken on the displacements and not on the phase field variable, hence the observation matrix is non-zero only in the part related to the displacement vector. The measurement data can be collected on arbitrary points throughout the domain which are not restricted to the nodes of the FE mesh. Finally, sensor noise is modelled with the Gaussian $\bm{\eta}_n$ whereas the model-data-mismatch is captured with another Gaussian $\bm{\delta}_n$. Note that $\bm{\delta}_n,\bm{\eta}_n$ are independent of $\bm{a}_n$. To do so, a kernel function, e.g. of squared exponential or Matérn type, has to be chosen and this kernel is then evaluated at the sensor points. In the numerical examples given in this manuscript, the Matérn kernel
\begin{equation}
     c(\bm{x},\bm{x}') = \sigma^2 \frac{2^{1-\nu}}{\Gamma(\nu)}  \left( \frac{\sqrt{2\nu}\| \bm{x} - \bm{x}' \|}{l}  \right)^{\nu} K_{\nu}  \left(  \frac{\sqrt{2\nu}\| \bm{x} - \bm{x}' \|}{l}   \right)
\label{eqn:MaternKernel}
\end{equation}
will be used throughout. 
It is defined with the scaling parameter $\sigma \in \mathbb{R}^+$, $\Gamma(\nu)$ the Gamma function and $K_{\nu}$ a modified Bessel function. Moreover, $l \in \mathbb{R}^+$ is the lengthscale parameter and $\nu\in \mathbb{R}^+$ controls the smoothness of the kernel.
For $\nu \to \infty$  (\ref{eqn:MaternKernel}) becomes the squared exponential kernel. See \cite{rasmussenGaussianProcessesMachine2006} for a thorough discussion.

The data model, connecting model to measurements, needs to be calibrated.
The calibration is accomplished by tuning the model error covariance. The corresponding hyperparameters $\bm{w} = (\nu, \sigma, l)$ can be learned from the given data by minimizing the negative log likelihood of the data model with an optimiser such as L-BFGS. A full sampling based approach, based on Markov Chain Monte Carlo (MCMC) can also be used, if uncertainties for the parameter estimation are sought.
With a calibrated data model, the Kalman filter equations, to be described below, can be derived by conditioning the FE solution fields on the data. For a single observation we obtain
\begin{equation}
\label{eqn:margLik}
p(\bm{y}\vert \bm{w}) = \frac{1}{(2\pi)^{n_{\text{sens}}/2}\lvert \bm{G} \rvert }  \exp \left(-\dfrac{1}{2} (\bm{y} - \rho \bm{H}\bar{\bm{a}} )^\top \bm{G}^{-1} (\bm{y} - \rho \bm{H} \bar{\bm{a}})\right)
\end{equation}
with $n_{\text{sens}}$ the number of sensors and
\begin{equation}
 \bm{G} = \rho^2 \bm{H} \bm{C}_{a} \bm{H}^\top  + \bm{C}_{\delta} + \bm{C}_{e}
\end{equation}
the covariance of \eqref{eqn:datamodel}.
Taking the natural logarithm yields
\begin{equation}
\label{eqn:logLikROM}
    \log p(\bm{y}\vert \bm{w}) = -\dfrac{n_{\text{sens}}}{2} \log (2 \pi) - \dfrac{1}{2} \log \lvert \bm{G} \rvert - \dfrac{1}{2} (\bm{y} - \rho \bm{H} \bar{\bm{a}})^\top \bm{G}^{-1} (\bm{y} - \rho \bm{H} \bar{\bm{a}} ) \;.
\end{equation}
%
For multiple observations per sensor we find
\begin{equation}
\log p(\bm{Y}\vert \bm{w}) = \sum_{i=1}^{n_\text{obs}} \log p(\bm{y}_i\vert \bm{w}) \;,
\end{equation}
which can be applied if at one sensor location ${n_\text{obs}}$ multiple independent observations exist. 
After optimising with regard to the given likelihood, the data model can be evaluated with the estimated hyperparameters.

In order to generate artificial measurement data, one can sample from the data model with a fixed set of parameters. Another possibility is misspecifying the model by changing parameters or adding complexity through different boundary terms etc. In this paper, a fixed initial phase field is chosen that isn't part of the prior ensemble. Also, in order to prevent an inverse crime \cite{kaipioStatisticalComputationalInverse2005}, the reference solution is computed on a different triangulation $\mathcal{T}_\text{data}$ with $h_\text{data}<h$ compared to the prior ensemble. An uncorrelated, mean-free noise term $\bm{e}\sim\mathcal{N}(0,\bm{C}_e)$ is added to the process.
The resulting data generating process is then sampled $n_\text{obs}$ times. 


    \subsection{Ensemble Kalman Filter}

    The EnKF is an equal-weight particle system variant of the Kalman filter \cite{evensenEnsembleKalmanFilter2003}.
    This enables using non-linear models without an explicit linearization. Another sampling-based variant is the particle filter \cite{doucetSequentialMonteCarlo2000}, where each ensemble member gets assigned a weight which is updated over time. Particle filters are known to work well for non-linear problems \cite{gilesParticleDAjlV10Realtime2024}, hence they appear promising in our regard. However, the problem with the particle filters is the scaling to higher dimensions, as shown in the literature \cite{bengtssonCurseofdimensionalityRevisitedCollapse2008}: Filter breakdown is expected at only hundreds of parameters already, but a high-dimensional state-space system, as the one used here, utilises tens of thousands or more state parameters. The named problems have also been addressed in the context of statFEM in \cite{duffinStatisticalFiniteElements2021}, where the authors chose an EnKF instead of the particle filter. Because of the limitations in regard to high-dimensional problems, we also resort to the EnKF in this paper. Within the EnKF, each ensemble member has the same weight and in the update step, as data comes in, each member is conditioned individually on data. 
    
    The EnKF consists of two repeating steps: First, in the prediction step, the approximate solution to the PDE is evolved in time. The resulting fields can then either directly be used as initial conditions to further evolve in time or be updated. Thus, the second step is the update, or analysis, step. If at some point in time measurement data are available, the solution fields of each ensemble member are updated via a Kalman shift, which can be framed in a Bayesian updating framework. 
    It should be noted that the EnKF converges to the true filter for an increasing particle size, only in the linear Gaussian setting \cite{calvelloStatisticalAccuracyEnsemble2025}. However, the phase field problem is inherently non-linear, hence the accuracy of our particle approximations is less obvious. Recent works develop theory for the near-Gaussian setting, see \cite{calvelloAccuracyEnsembleKalman2025,calvelloStatisticalAccuracyEnsemble2025}, and bounds for the Gaussian approximation, as discussed in \cite{sanz-alonsoInverseProblemsData2023a}, become available. 
    This introduction and also the implementation of the filter, presented below, follows \cite{duffinStatisticalFiniteElements2021,sanz-alonsoInverseProblemsData2023a}.

    It will be instructive to view the EnKF update from an optimization perspective. We also re-introduce the time index $n$ at this point. Following \cite{sanz-alonsoInverseProblemsData2023a}, we consider the objective function
    \begin{equation}
        J_{n,i}(\bm{a}^\text{A}_{n,i}) = \frac{1}{2} \left \| \bm{y}_{n} - \bm{H} \bm{a}^\text{A}_{n,i} \right \|_{\Gamma}^2 + \frac{1}{2} \left \| \bm{a}^\text{A}_{n,i} - \bm{a}^\text{F}_{n,i} \right \|_{\bm C_{n}}^2,
        \label{eqn:optimEnKF}
    \end{equation}
where the analysis vector $\bm{a}^\text{A}_{n,i}$, for each ensemble member $i$, is determined by minimization. The forecast $\bm{a}^\text{F}_{n,i}$ is obtained by propagating the particles according to the phase field model from the previous time step, to be detailed in Section \ref{subsec:prediction}. For simplicity, only a single time step $n$ and the corresponding data are considered here. We can observe that the model is not strongly enforced, hence the analysis can potentially violate constraints that would be imposed be the phase field model; unless $\bm C_{n}$ tends to zero and the data are without any influence. Because the non-linear model is not part of the optimization, a closed form can be derived, i.e. \eqref{eqn:EnKFUpdate}. At the same time, we observed that the standard EnKF update can fail and we attribute this precisely to the missing explicit influence of the phase field model in the update equations, in combination with the strong nonlinearities of the phase field equations. In order to resolve this issue, a constraint approach will be introduced below.

    The EnKF has to be initialised at first. As it is a sampling based approach, the forward problem is solved for $n_{\text{ens}}$ samples of the input parameters. Uncertain input parameters could be the spatial configuration of an initial damage field, a random material parameter or probabilistic forcing with random perturbations in the initialisation or at each timestep. 
    
    \subsubsection{Forecast}
    \label{subsec:prediction}
    Consider the ensemble $\{\bm{a}_{{n-1},i}\}_i$ at time/loading step $n-1$, originating from either the initialization of the filter or a previous prediction or update. Note that we omit the dependence on $\bm{y}_{1:n-1},\bm{w}_{1:n-1}$.
    %
    For each ensemble member, the prediction $\bm{a}^{\text{F}}_{n,i} = (\bm{a}^{\text{F,u}}_{n,i}, \bm{a}^{\text{F,d}}_{n,i})$ is computed from the residual equation system \eqref{eq:discrete_residual_n}, 
%
    %
    Subsequently, the prediction's mean and covariance can be computed as
    \begin{equation}
         \bar{\bm{a}}_{n}^{\text{F}} = \frac{1}{n_{\text{ens}}} \sum_{i=1}^{n_{\text{ens}}} \bm{a}_{n,i}^{\text{F}}
    \end{equation}
    and
    \begin{equation}
        \bm{C}_{n}^{\text{F}} = \frac{1}{n_{\text{ens}}-1} \sum_{i=1}^{n_{\text{ens}}} \left( \bm{a}^{\text{F}}_{n,i} - \bar{\bm{a}}_{n}^{\text{F}} \right)  \left( \bm{a}_{n,i}^{\text{F}} - \bar{\bm{a}}_{n}^{\text{F}} \right)^{\top},
    \end{equation}
    respectively. To keep the notation short, from here on we drop the dependence of the forecast on previous time steps, as e.g. in $\bar{\bm{a}}_{n|n-1}^{\text{F}}$. The prediction step is iteratively executed until reaching a time step at which data are available. Then, each ensemble member is updated with respect to the given measurement data. In the 1D numerical example, data are available from time step $82$ on in an interval of $10$ steps. Because a finite ensemble size is used, the covariance is usually inflated and a localization procedure is employed. Details on that can be found in \ref{app1}.



    \subsubsection{Analysis}

    The first step towards updating the prior state, i.e. conditioning it on new data $\bm{y}_n$, is updating the hyperparameters $\bm{w}_{1:n-1}$ of the data model, in case that they change over time. If so, the negative log-marginal posterior of the data model is minimised given suitable priors for the individual parameters. With the updated hyperparameters $\bm{w}_{1:n}$, each ensemble member is updated by a Kalman shift. The Kalman shift is an analytical solution to the optimization problem \eqref{eqn:optimEnKF}, see \cite{sanz-alonsoInverseProblemsData2023a} for details. For the state of one member and a single observation per sensor we obtain
                \begin{equation}
           \begin{split}
               &\bm{a}^{\text{A}}_{n,i} = \bm{a}^{\text{F}}_{n,i} + \bm{C}^{\text{F}}_{n} \bm{H}^{\top} \left(\rho^2 \bm{H} \bm{C}^{\text{F}}_{n}\bm{H}^{\top} + \bm{C}_{\delta} + \bm{C}_{e} \right)^{-1}\left(\bm{y}_{n} -\rho \bm{H} \bm{a}^{\text{F}}_{n,i}\right).
           \end{split}
    \end{equation}
%

    For the data model as proposed in \cite{girolamiStatisticalFiniteElement2021}, i.e. with multiple available readings per sensor, there holds for the updated state of a single ensemble member

       \begin{equation}
       \label{eqn:EnKFUpdate}
        \bm{a}^{\text{A}}_{n,i} = \bm{a}^{\text{F}}_{n,i} + \bm{C}^{\text{F}}_{n} \bm{H}^{\top} \left(\rho^2 n_{\text{obs}} \bm{H} \bm{C}^{\text{F}}_{n}\bm{H}^{\top} + \bm{C}_{\delta} + \bm{C}_{e} \right)^{-1} \left( \sum_{j=1}^{n_{\text{obs}}}\bm{y}_{n,j} -\rho n_{\text{obs}} \bm{H} \bm{a}^{\text{F}}_{n,i}\right).
    \end{equation}

\begin{figure}[ht!]
    \centering
        \includegraphics[width=12cm]{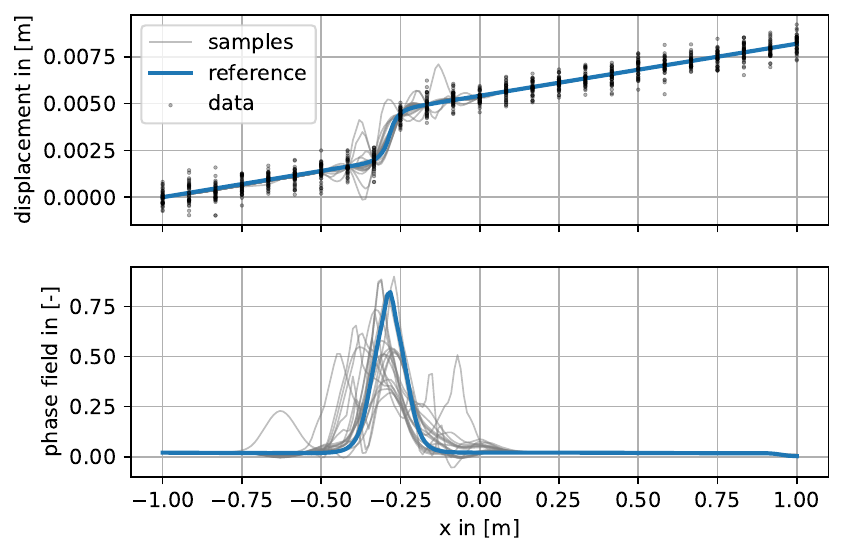}
    \caption{The posterior ensemble in gray compared to the reference solution, from which noisy measurements are drawn, in blue. Compared to the prior, the ensemble now clusters more closely around the reference. However, numerical artefacts and unphysical phase field solutions appear, which can lead to the next time step not converging. This posterior ensemble, or analysis, is the result of row $10$ in Algorithm \ref{alg:kalman}.}
    \label{fig:ShiftNoReg}
\end{figure}

    The updated states now serve as an initial condition again for the next prediction step. In Figure \ref{fig:ShiftNoReg}, the ensemble analysis for the running example are visible. The ensemble members now cluster closely around the reference solution, although it is clearly visible that the posterior ensemble members exhibit strong unphysical noise. Also, the phase field drops below zero, which violates the constraints encoded in the system of PDEs. Subsequent Newton iterations for the next time step might not converge due to this effect. The reason for these problems is the strong gradient in the displacement field.
The problem can be overcome to some extent by smoothing the displacement field before the update, as visible in Figure \ref{fig:regularisationChart1} a). The resulting posterior ensemble exhibits much less noise, yet negative phase fields still occur and the displacements are still unphysical. This unphysical behaviour of the posterior can be overcome by regularization within a staggered solution of the individual equations for phase field and displacement. This will be discussed in the next section.
%

In Algorithm \ref{alg:kalman}, the forecast time stepping as well as the EnKF analysis are summarized.
$\bm{T}$ denotes the set of pseudo-time steps at which the phase field problem is solved with different loadings. 
$\bm{T}_\text{D}$ denotes the set of time steps at which data is available and the EnKF update is performed.
\begin{algorithm}[ht!]
\caption{Phase Field Ensemble Kalman Filter}\label{alg:kalman}
\begin{algorithmic}[1]
\Require $\bm{a}_0^{\text{F}} =  \left\{(\bm{a}_{0,i}^{\text{F},u},\bm{a}^{\text{F},d}_{0,i})\right\}_{i=1}
^{n_{\text{ens}}}$, $t=t_0 = 0$
\Comment{Initial ensemble at $t=0$}

\For{$t \in \bm{T}$}
\Comment{Calculate next pseudo-time forecast...}
 \State $t = t + \Delta t$
 \State $n = n+1$
 \For{$i = 1$ to $n_{\text{ens}}$}
 \Comment{...for each ensemble member}
     \State update $\check{\bm{a}}^d, \bm{\varphi}_{n-1}$ based on $\bm{a}^\text{F}_{n-1,i}$
     \State solve monolithically:\\
     $\bm{R}^u(\bm{a}^u, \check{\bm{a}}^d; \bm{\varphi}_{n-1}, \ell) = 0$,\\
    $\bm{R}^d(\bm{a}^u, \bm{a}^d; \bm{\varphi}_{n-1}, \ell) = 0$
     \State set $\bm{a}^\text{F}_{n,i}$ equal to $(\bm{a}^u, \bm{a}^d)$
  \EndFor
\State $\bm{a}_n^{\text{F}} =  \left\{(\bm{a}_{n,i}^{\text{F},u},\bm{a}^{\text{F},d}_{n,i})\right\}_{i=1}
^{n_{\text{ens}}}$
  \If{$t \in \bm{T}_\text{D}$}
        \For{$i = 1$ to $n_{\text{ens}}$}
     \State perform EnKF analysis Eq. \eqref{eqn:EnKFUpdate} $\bm{a}^\text{F}_{n,i} \rightarrow \bm{a}^\text{A}_{n,i}$
  \EndFor
  \State $\bm{a}_n^{\text{A}} =  \left\{(\bm{a}_{n,i}^{\text{A},u},\bm{a}^{\text{A},d}_{n,i})\right\}_{i=1}
^{n_{\text{ens}}}$
          \For{$i = 1$ to $n_{\text{ens}}$}
     \State perform iterative regularization (following section) $\bm{a}^\text{A}_{n,i} \rightarrow \bm{a}^\text{R}_{n,i}$
  \EndFor
  \State $\bm{a}_n^{\text{R}} =  \left\{(\bm{a}_{n,i}^{\text{R},u},\bm{a}^{\text{R},d}_{n,i})\right\}_{i=1}
^{n_{\text{ens}}}$
  \State $\bm{a}_n^\text{F} = \bm{a}_n^\text{R}$
  \EndIf
\EndFor

\end{algorithmic}
\end{algorithm}

\subsubsection{Regularization}
\label{sec:reg}
Figure \ref{fig:ShiftNoReg} depicts the updated ensemble for the running example. It is clearly visible that the Kalman update can generate both spurious oscillations in the displacements as well as phase field variables with negative values. Several paths can be pursued to remove these effects. One could transform the phase field variable to enforce the bounds $[0,1]$ and smooth the displacements, however, we found that in particular the smoothing is difficult to realize in higher dimensions. Another option is to enforce the phase field model as a strong constraint in a 3D-Var or 4D-Var type approach. This however, appears to be very costly to realize and is left for future research.

Here, we propose a different solution, where we regularize the Kalman shift with a few well-chosen staggered iterations using the phase field model. 
Instead of a standard Gauss Newton approach to solve the individual equations, we treat each Newton step as a single-step approximation to a proximal operator, which serves to balance between the purely PDE-based regularisation and the data influence from the EnKF analysis. The reasoning behind this will be discussed in Section \ref{sec:discussion}.

The first step towards regularizing the posterior is to compute the displacement and phase field again at the same pseudo-time step. Instead of the monolithic approach used in the prediction, we solve for $\bm{u}$ and $d$ individually and with a larger phase field length scale $L > \ell$. 

\begin{enumerate}
\item Given the result from the EnKF update, i.e. $\bm{a}_{n,i}^{\text{A},u}$, solve
\begin{equation}
\bm{R}^d(\bm{a}^{\text{A},u}_{n,i}, \tilde{\bm{a}}^{\text{R},d}_{n,i}; \bm{0}, L) = 0
\end{equation}
for $\tilde{\bm{a}}^{\text{R},d}_{n,i}$.

%
\item Solve for $\tilde{\bm{a}}^{\text{R},u}_{n,i}$ subject to
\begin{equation}
\bm{R}^u(\tilde{\bm{a}}^{\text{R},u}_{n,i}, \tilde{\bm{a}}^{\text{R},d}_{n,i}; \bm{0}, L) = 0.
\end{equation}
Note that, with $\bm{\varphi}=0$, we essentially removed the irreversibility constraint at this step. The notation $\bm{F}_{1;n}$ reflects, that we are still imposing the prescribed displacement at step $n$. 
\end{enumerate}

The result is visible in Figure \ref{fig:regularisationChart1} b). The phase field appears much wider, while still pinning the reference solution accurately. The wider phase field covers most of the noise around the position of the crack. This is also reflected in the corresponding displacement, computed with the regularized phase field. A downside is that although the displacement indeed pins the correct reference position of the crack, the remaining stiffness at the crack is lower than in the reference solution. The reason for that is the increase in the overall, cumulated, damage in the domain due to the wider phase field. In turn, that leads to a faster propagation speed of the crack in higher dimensional settings. Once a crack is fully developed, the effect is not relevant any more. Nevertheless, we consider this effect to be preferable to the noisy, unphysical posterior. 

The second part consists in recomputing the phase field from the regularized displacement, this time with the original length scale:
\begin{enumerate}
\setcounter{enumi}{2}
\item Given the regularized displacements, i.e. $\tilde{\bm{a}}^{\text{R},u}_{n,i}$, find ${\bm{a}}^{\text{R},d}_{n,i}$ subject to 
\begin{equation}
    \bm{R}^d(\tilde{\bm{a}}^{\text{R},u}_{n,i},\bm{a}^{\text{R},d}_{n,i}; \bm{0}, \ell) = 0.
\end{equation}

\item 
Using the phase field model without the reversibility constraint 
find $\bm{a}^{\text{R},u}_{n,i}$ such that
\begin{equation}
    \bm{R}^u(\bm{a}^{\text{R},u}_{n,i}, \bm{a}^{\text{R},d}_{n,i}; \bm{0}, \ell) = 0.
\end{equation}

\end{enumerate}
The result is visible in Figure \ref{fig:regularisationChart1} c). 
Clearly, there is still some uncertainty left, which is expected due to a limited amount of sensors and measurement noise.
For comparison, the regularization scheme is applied to a completely fractured 1D domain, i.e. with a phase field maximum of $1$. These results are visible in Figure \ref{fig:FullCrackStagger}.



\begin{figure}
\centering
\subfloat[]{\includegraphics[width=0.9\columnwidth]{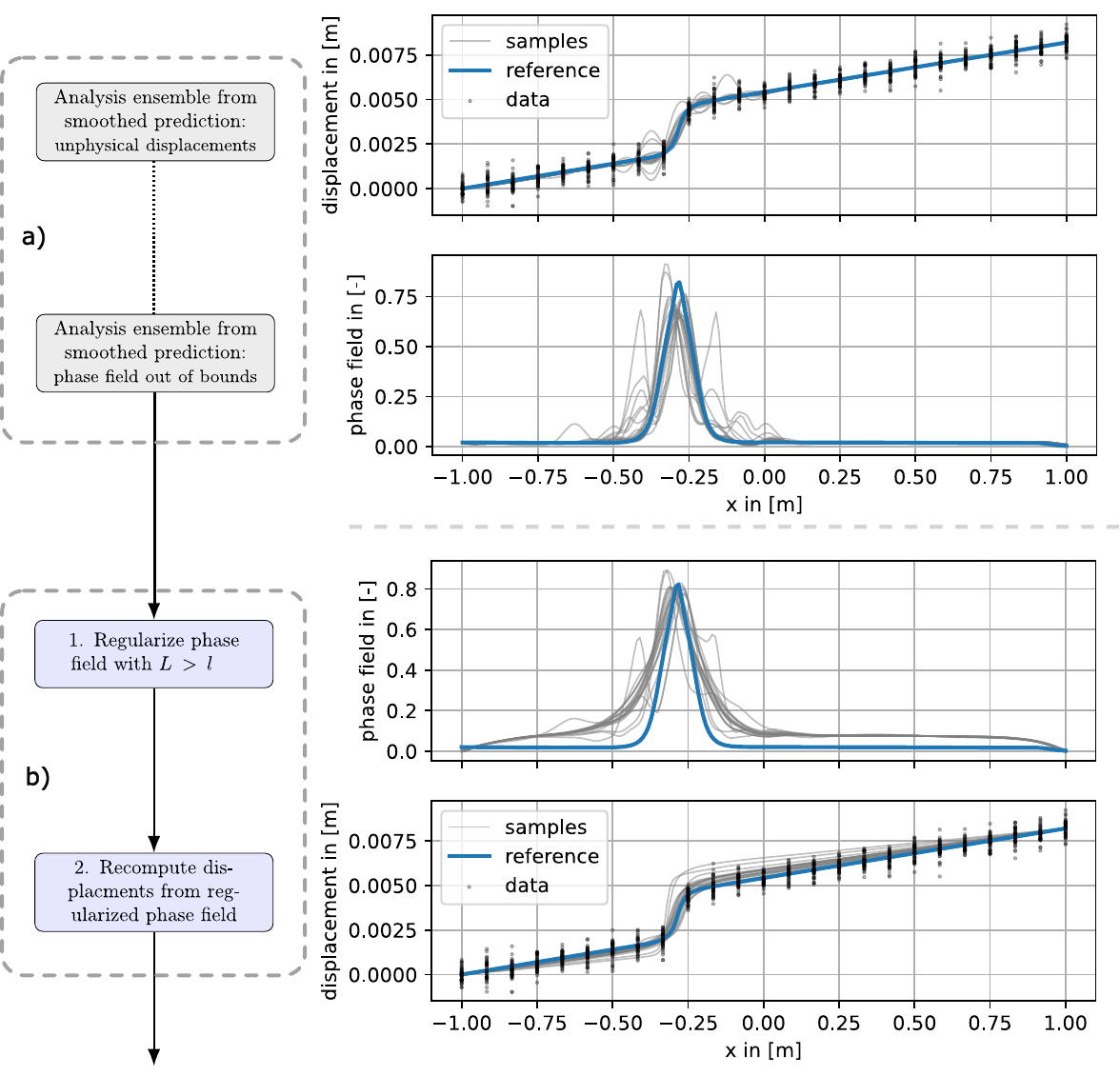} \, 
}
\caption{a): Posterior ensemble with prior smoothing through a Gaussian Process. The strongest artifacts are filtered out, but the unphysical phase field remains. \\
b): The posterior ensemble in gray after the first regularization step compared to the reference solution in blue. By the projection on a solution space with larger phase field length scale $L$, most of the numerical artifacts are filtered. However, the larger length scale leads to more accumulated damage in the whole domain and therefore a loss in total stiffness, which reflects in a slightly larger displacement. This intermediate result corresponds to the results of rows $2-5$ in Algorithm \ref{alg:stagger}.
}
\label{fig:regularisationChart1}
\end{figure}

\begin{figure}
\ContinuedFloat
\centering
\subfloat[]{\includegraphics[width=0.9\columnwidth]{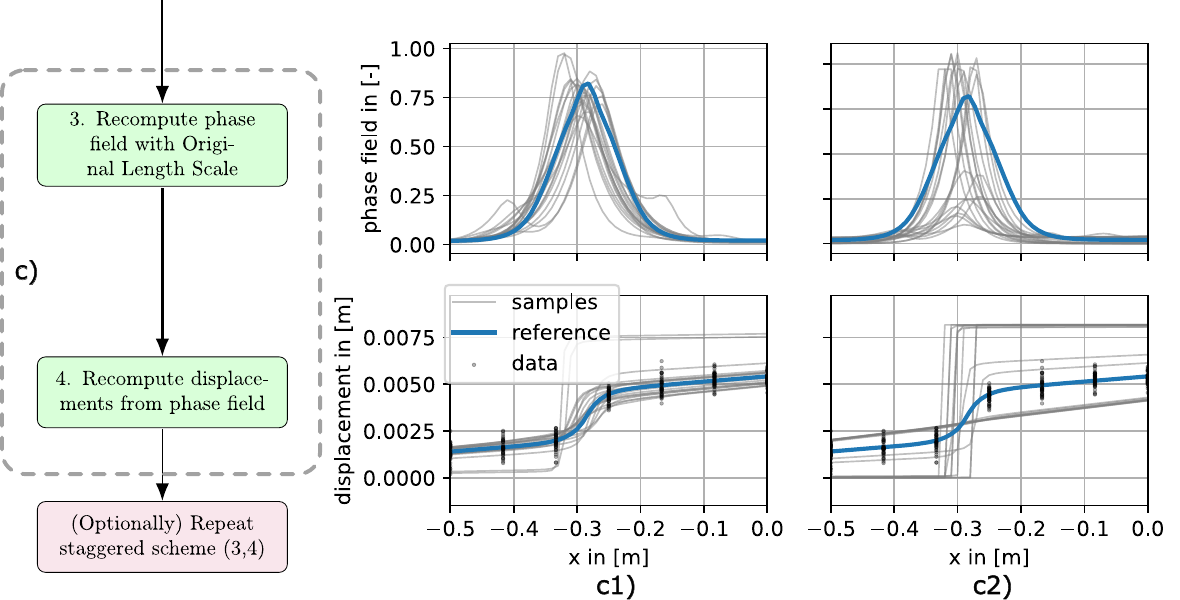} \, 
\label{fig:regularisationChart2}}%
\caption{c): Correction with the original phase field length scale. On the left in c1), the process has been repeated once. On the right in c2), the process has been repeated four times. It is visible that a single repetition yields results closer to the reference. This is because inference takes place on two phenomena here: fracture initiation and fracture localization. Bayesian treatment of initiation appears much harder in this context; the iterations lead to results which indicate mostly complete damage or only very little damage. However, fracture localization works as desired: the correct position of the crack is found. We are mostly interested in fracture localization at this point. This regularized ensemble corresponds to row $13$ in Algorithm \ref{alg:stagger}.}
\label{int}
\end{figure}

\begin{figure}[ht!]
    \centering
    \includegraphics[width=11cm]{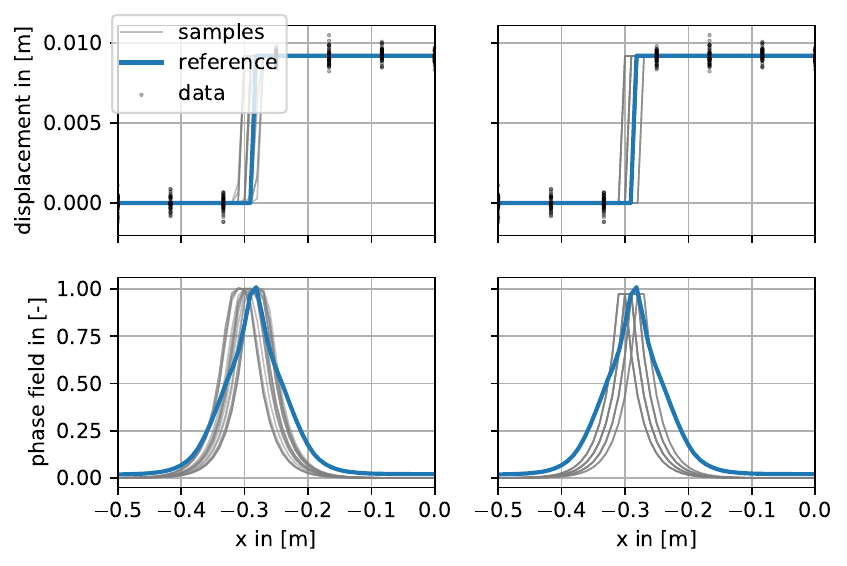}
    \caption{Here, the EnKF analysis and regularization procedure have been performed at a later time step, where the reference solution and all ensemble members already indicate complete damage at a respective position. On the left, the process has been repeated once. On the right, the process has been repeated four times. It is visible that a single repetition yields good results already. Repeating the staggered scheme leads to similar results, but with converged ensemble member phase fields shapes. Crack nucleation does not play a role here, hence inference is simpler and the analysis exhibits a smaller standard deviation in the samples. }
    \label{fig:FullCrackStagger}
\end{figure}

Finally, the full state vector is reconstructed as
\begin{equation}
\bm{a}_n^{\text{R}} =  \left\{(\bm{a}_{n,i}^{\text{R},u},\bm{a}^{\text{R},d}_{n,i})\right\}_{i=1}
^{n_{\text{ens}}}
\end{equation}
and serves as start value for the next pseudo time step. Algorithm \ref{alg:stagger} summarizes the described regularization procedure.

\begin{algorithm}[ht!]
\caption{Regularizing the Kalman analysis }\label{alg:stagger}
\begin{algorithmic}[1]
\Require $\bm{a}_n^{\text{A}} =  \left\{(\bm{a}_{n,i}^{\text{A},u},\bm{a}^{\text{A},d}_{n,i})\right\}_{i=1}
^{n_{\text{ens}}}$
\Comment{Ensemble after EnKF analysis}
\Ensure $L > \ell$
\For{$i=1$ to $n_{\text{ens}}$}
\Comment{Regularize each ensemble member}
\State with $\bm{a}_{n,i}^{\text{A},u}$ and
$\bm{\varphi}_{n-1} \stackrel{!}{=} 0$
\State {$\quad$} solve $\bm{R}^d(\bm{a}^{\text{A},u}_{n,i}, \tilde{\bm{a}}^{\text{R},d}_{n,i}; \bm{0}, L) = 0$ 
for $\tilde{\bm{a}}^{\text{R},d}_{n,i}$
\State with $\tilde{\bm{a}}^{\text{R},d}_{n,i}$ and $\bm{\varphi}_{n-1} \stackrel{!}{=} 0$
\Comment{Degrade stresses}
\State {$\quad$} solve 
$\bm{R}^u(\tilde{\bm{a}}^{\text{R},u}_{n,i}, \tilde{\bm{a}}^{\text{R},d}_{n,i}; \bm{0}, L) = 0$
%
for $\tilde{\bm{a}}^{\text{R},u}_{n,i}$
\State with $\tilde{\bm{a}}^{\text{R},u}_{n,i}$ and
$\bm{\varphi}_{n-1} \stackrel{!}{=} 0$
\State {$\quad$} solve 
$\bm{R}^d(\tilde{\bm{a}}^{\text{R},u}_{n,i},\bm{a}^{\text{R},d}_{n,i}; \bm{0}, \ell) = 0$
%
%
%
for $\bm{a}^{\text{R},d}_{n,i}$
\For{$j=1$ to $n_{\text{stagger}}$}
\Comment{Staggered solution scheme}
\State with $\bm{a}^{\text{R},d}_{n,i}$ and $\bm{\varphi}_{n-1} \stackrel{!}{=} 0$
\State {$\quad$} solve $\bm{R}^u(\bm{a}^{\text{R},u}_{n,i}, \bm{a}^{\text{R},d}_{n,i}; \bm{0}, \ell) = 0$ for $\bm{a}^{\text{R},u}_{n,i}$
\State with $\bm{a}^{\text{R},u}_{n,i}$ and
$\bm{\varphi}_{n-1} \stackrel{!}{=} 0$
\State {$\quad$} solve $\bm{R}^d({\bm{a}}^{\text{R},u}_{n,i},\bm{a}^{\text{R},d}_{n,i}; \bm{0}, \ell) = 0$ for $\bm{a}^{\text{R},d}_{n,i}$
\EndFor
\EndFor

\State $\bm{a}_n^{\text{R}} =  \left\{(\bm{a}_{n,i}^{\text{R},u},\bm{a}^{\text{R},d}_{n,i})\right\}_{i=1}
^{n_{\text{ens}}}$
\Comment{Ensemble after regularization}

\end{algorithmic}
\end{algorithm}

The regularization process is similar to the staggered approach towards solving a phase field problem \cite{kirkesaetherbrunIterativeStaggeredScheme2020,hofackerContinuumPhaseField2012} in the first place. If subsequent Newton iterations still fail to converge, the staggered scheme can be repeated. It is worth noting, however, that the stiffness can either decrease or increase further with every new iteration of the staggered scheme and hence the crack propagation speed either increases or decreases, as it can be noticed in Figure \ref{fig:regularisationChart1} c2). The inferred position of the crack will still be inferred correctly. This effect highlights that the proposed method is able to successfully perform crack localization, which is the desired functionality for higher dimensional settings, but when repeating the staggered scheme the method is less successful at accurately inferring the magnitude of the phase field at the crack front. 

\begin{figure}[ht!]
    \centering
    \includegraphics[width=16.0cm]{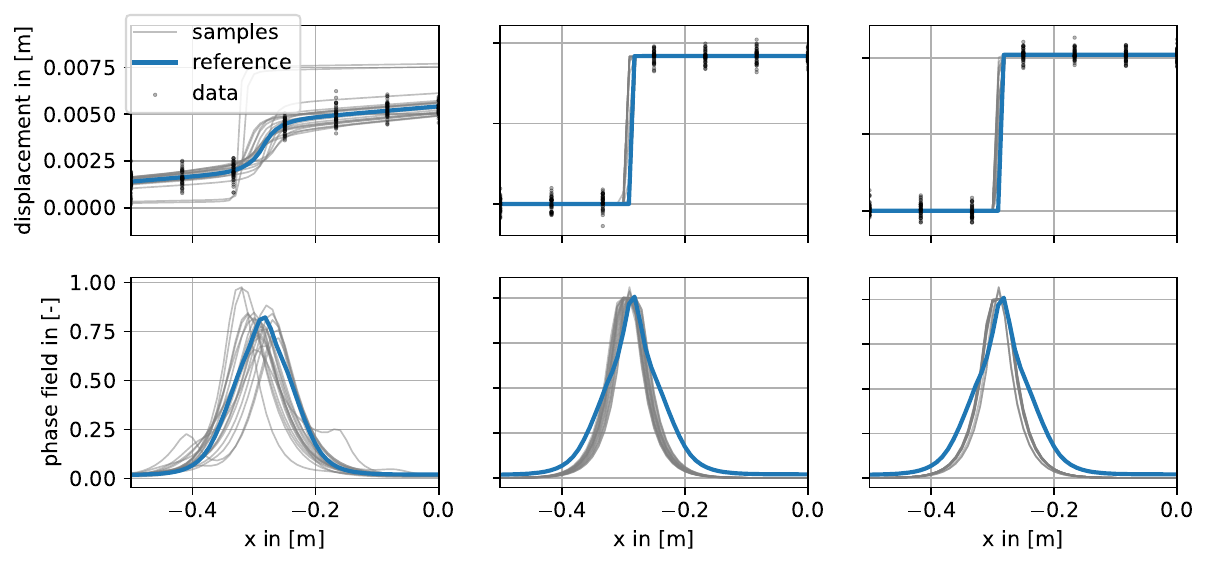}
    \caption{The more EnKF analysis steps are included at different time steps (82, 92 and 102), the smaller the uncertainty about the crack position becomes: Displacements and phase field converge towards the reference. At each time step, the staggered scheme has been repeated once.}
    \label{fig:multipleUpdates1D}
\end{figure}

After regularization, the next pseudo-time prediction can be computed. Figure \ref{fig:multipleUpdates1D} shows more snapshots of the analysis timeline: With more updates, the ensemble clusters more closely around the reference, hence the uncertainty decreases.

\begin{figure}[ht!]
    \centering
    \includegraphics[width=16.0cm]{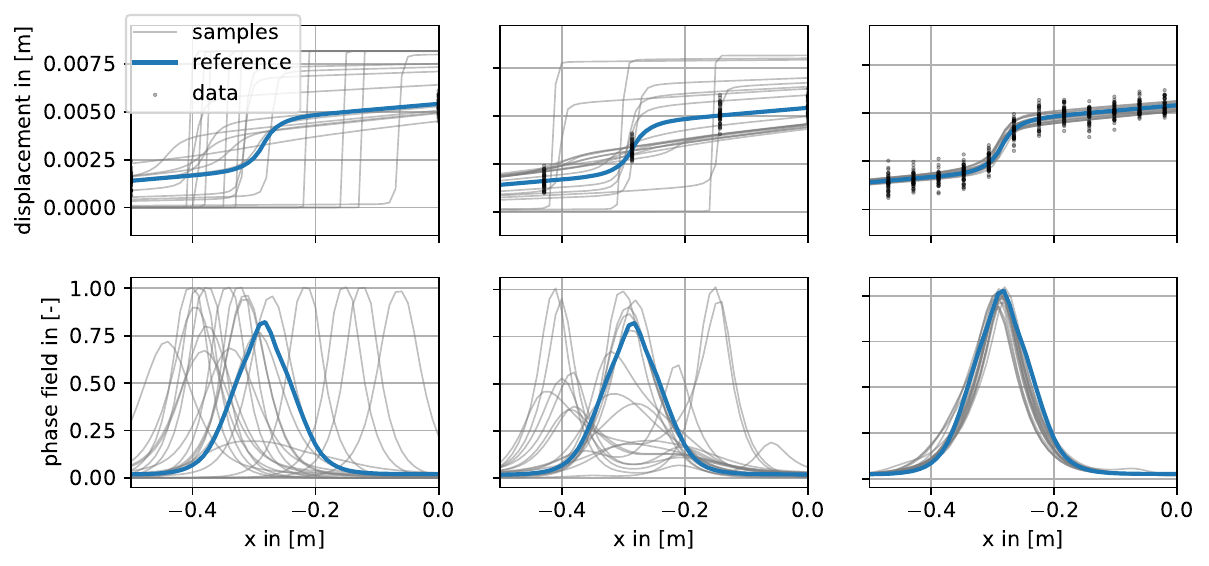}
    \caption{The more sensors positions (5, 15, 50) are used in the data generation and analysis, the smaller the uncertainty about the crack position becomes. Compare to Figure \ref{fig:multipleUpdates1D} with 25 sensor positions. All plots show the first analysis at time step 82.}
    \label{fig:SensorCount}
\end{figure}

The number of sensors has an influence on the quality of the posterior ensemble. Intuitively, more sensors imply a larger weight on the data and less on the prior. Figure \ref{fig:SensorCount} shows the analysis at the same time step, but with different numbers of sensors in the domain. 

\section{Numerical Study}
\subsection{Single Edge Notched specimen under Shear (SENS)}

\begin{figure}[ht!]
    \begin{minipage}[b]{.55\linewidth}
    \centering
    \includegraphics[page=1]{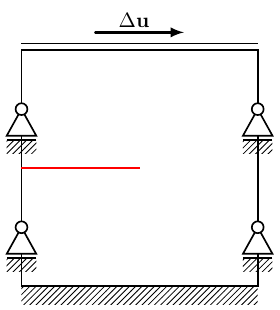}
    \captionof{figure}{SENS numerical experiment. Edge length: 1.0mm,\\ notch length: 0.5mm.}
        \label{fig:SENSmech}
  \end{minipage}\hfill
  \begin{minipage}[b]{.45\linewidth}
    \centering
\begin{tabular}{@{}ll@{}}
\toprule
\textbf{Parameter} & \textbf{Value}      \\ \midrule
Fracture Model & AT2                 \\
Energy Split   & Volumetric-Deviatoric   \\
$E$            & $210000$ {[}N/$\text{mm}^2${]} \\
$\nu$          & $0.3$ {[}-{]}       \\
$G_\text{c}$          & $2.7$ {[}N/mm{]}    \\
$l$            & $1.5e-2$ {[}mm{]}   \\
$\alpha$       & $\beta G_c /l$      \\ \bottomrule
\end{tabular}
    \captionof{table}{Parameters for the phase field model of the 2D numerical example}
  \end{minipage}
\end{figure}

For a 2D example, the SENS benchmark is chosen. As visible in Figure \ref{fig:SENSmech}, the mechanical setup consists of a unit square in [mm], fixed at the bottom. No vertical displacement is permitted at the left and right edges. A notch, depicted in red, of half of the edge length is modelled within the FEM mesh. The structure is loaded in a quasi-static manner at the top boundary. The Dirichlet boundary condition on the top boundary is used to prescribe a displacement increment of $\Delta u = 1e-4$ [mm] per pseudo time step for the first $70$ pseudo time steps. The increment decreases to $\Delta u = 1e-5$ [mm] from that time step on.

The ground truth, i.e. the solution fields from which noisy data are measured, is visible in Figure \ref{fig:GT2D}. Data of the displacement in both spatial dimensions is collected at sparse sensor locations ($n_\text{sens}<< n_\text{dof}$), denoted with a black cross. 
The parameters for the data assimilation are visible in Table \ref{tab:dataAssim2D}.
\begin{figure}[ht!]

    \begin{minipage}[b]{.35\linewidth}
    \centering
    
\begin{tabular}{@{}ll@{}}
\toprule
\textbf{Parameter}  & \textbf{Value} \\ \midrule
$n_{\text{ens}}$           & $100$          \\
$n_{\text{sens}}$          & $100$          \\
$n_{\text{obs}}$           & $20$           \\
$\sigma_e$          & $4e-4$         \\
$n_{\text{dof}}$ &   $34071$             \\
$n_{\text{dof}}$ ground truth & $36070$               \\
$r$                 & $1.05$         \\
$l_{\text{loc}}$           & $0.45$         \\ \bottomrule
\end{tabular}
    \captionof{table}{Parameters for the data assimilation problem}
    \label{tab:dataAssim2D}

  \end{minipage}\hfill
  \begin{minipage}[b]{.6\linewidth}
    \centering
\begin{tabular}{@{}ll@{}}
\toprule
\textbf{Parameter}  & \textbf{Value} \\ \midrule
$l_{\text{reg}}$           & $4l$          \\
$n_{\text{iter}}$          & $4$          \\ \bottomrule
\end{tabular}
    \captionof{table}{Regularization parameters}
    \label{tab:RegPars2D}
  \end{minipage}
\end{figure}
\begin{figure}[ht!]
    \centering
    \includegraphics[]{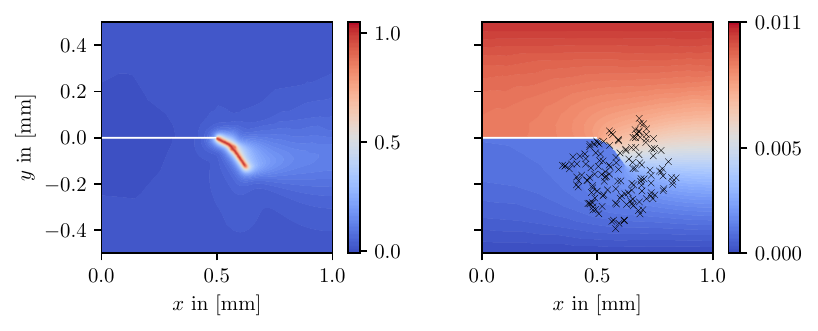}
    \caption{Data are collected from the displacement field of the depicted phase field solution. The EnKF is supposed to restore the phase field and displacements from a prior process and the measurements only. The sensor locations are marked with black crosses.}
    \label{fig:GT2D}
\end{figure}
The prior ensemble is constructed by randomly sampling $n_{\text{ens}}$ different initial conditions for the phase field. Each sample is defined by a single pore in the x-y plane in proximity to the slit. The position of the pore is defined with two random variables
\begin{align}
    X_0 &\sim 0.51+ (0.62-0.51) \cdot\text{Beta}(8,8)\\
    Y_0 &\sim -0.11+ (0.02-(-0.11)) \cdot\text{Beta}(8,8).
\end{align}

\begin{figure}[ht!]
    \centering
    \includegraphics[width=16cm]{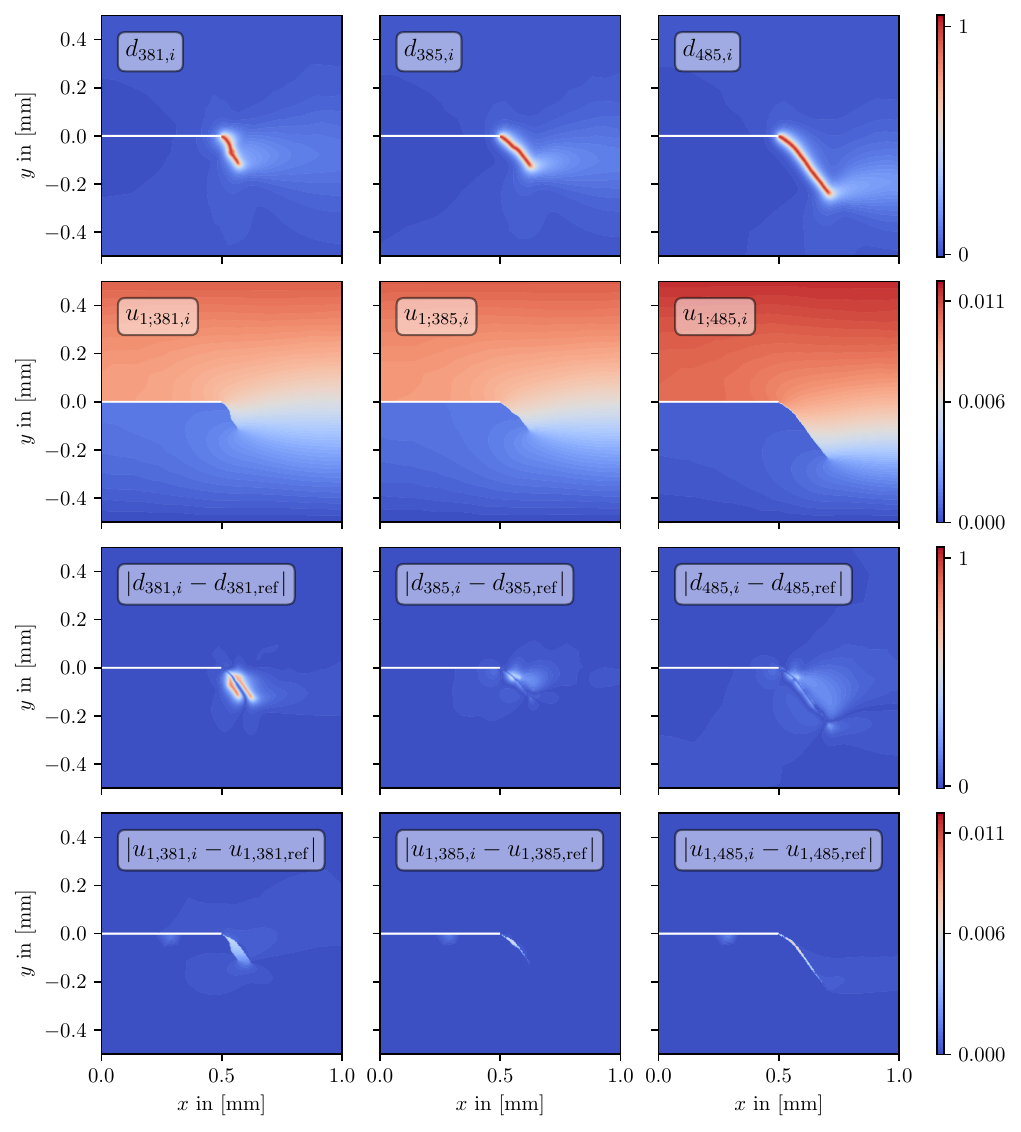}
    \caption{SENS analysis timeline for a single ensemble member. From left to right, three different time steps are visible. At time step $381$, no data assimilation has been performed yet. An error both in phase field and displacement is noticeable. At time step $382$, the first data assimilation and regularization procedure has been performed. As the regularization introduced smoothing, a couple of time steps are necessary for the sample to obtain a sharp crack again. At time step $385$, it is hence now visible that the sample changed and resembles the ground truth more closely. The error decreased. At time step $485$, a second analysis has been performed. The error is still low.}
    \label{fig:stateUpdates}
\end{figure}

In Figure \ref{fig:stateUpdates}, a single prior sample of displacement in $x$-direction and the corresponding phase field is visible along with the absolute error. The second column shows the timestep after an EnKF analysis step. Qualitatively, the absolute error reduced, i.e. the posterior phase field aligns more closely with the ground truth. The third column shows a later time step after a second analysis. The error remains small.
The chosen ensemble member exhibits a particularly strong deviation from the ground truth. Still, the analysis leads to a small error. Notice that this is also the case for all other ensemble members, i.e. the error either decreases or stays low if there is already only a small deviation.

\begin{figure}[ht!]
    \centering
    \includegraphics[width=16cm]{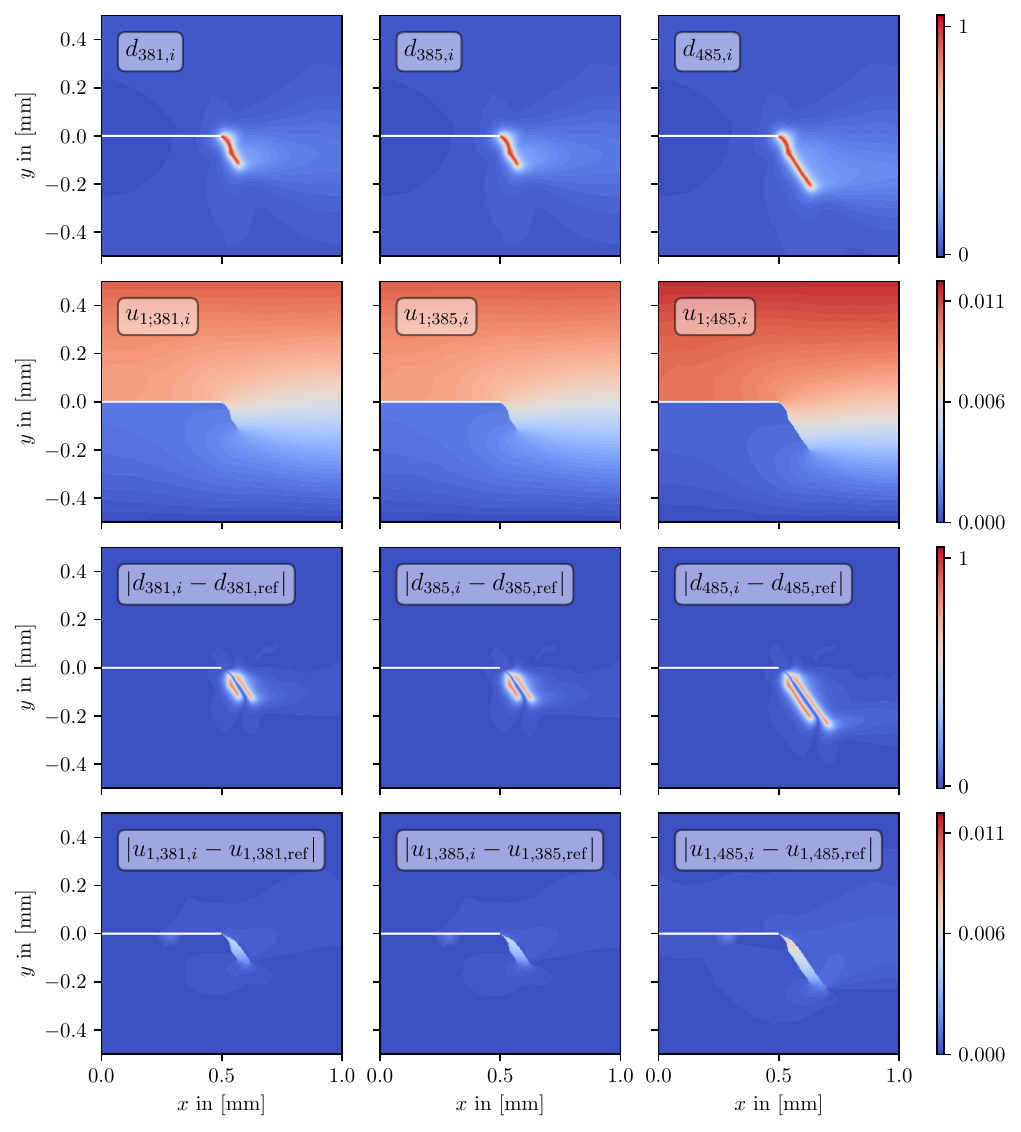}
    \caption{SENS timeline for a single ensemble member, without data assimilation. It is visible that the error increases with every time step.}
    \label{fig:stateUpdatesNOUP}
\end{figure}

In Figure \ref{fig:stateUpdatesNOUP}, on the other hand, no analysis steps have been applied to the sample and hence the error increases over time, both in displacements and phase field. Hence, it can be stated that performing a Kalman shift on the prior ensemble indeed leads to a reduction in the error of the individual sample and therefore also in the uncertainty of the whole ensemble.

\subsubsection{Reaction forces and extrapolation after the update}
\begin{figure}[ht!]
    \centering
    \includegraphics[width=10cm]{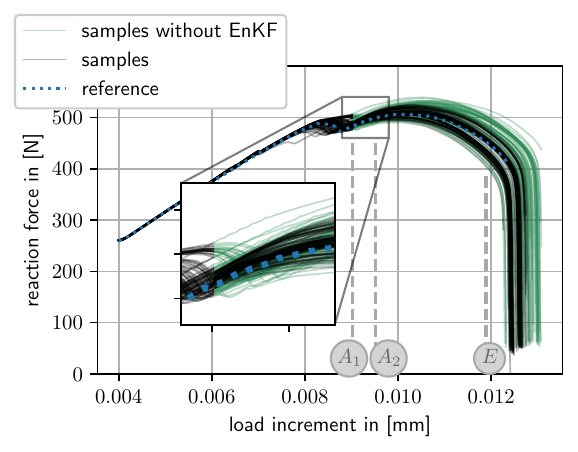}
    \caption{Reaction forces for the SENS benchmark with a single analysis step at time step $A_1$, visible in black. After the Kalman shift, the ensemble members cluster more closely around the reference solution. For comparison, the ensemble without Kalman shift is visible in green.
    }
    \label{fig:2dreact}
\end{figure}

\begin{figure}[ht!]
    \begin{minipage}[b]{.45\linewidth}
    \centering
     \includegraphics[]{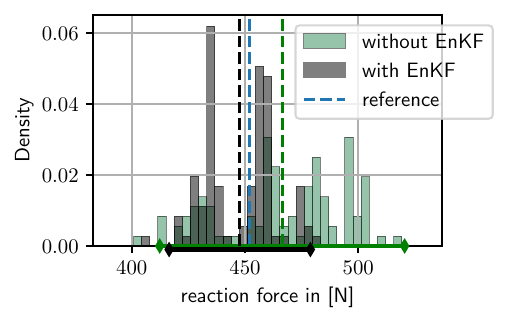}
    \caption{Histogram of the distribution of reaction forces at extrapolation time step $E$ for the SENS benchmark with a single analysis step at time step $A$. The spread ($2\sigma$) reduced while the reference solution still lies in the range of the the posterior covariance.}
    \label{fig:2dreactKDE}
  \end{minipage}\hfill
  \begin{minipage}[b]{.45\linewidth}
    \centering
\includegraphics[]{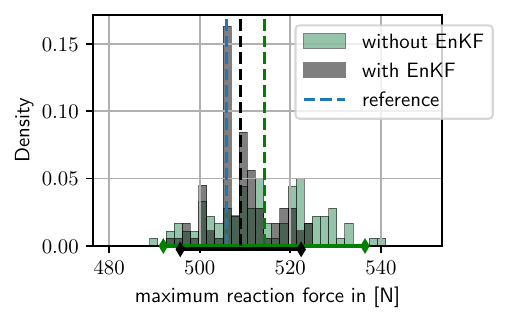}
    \caption{Histogram of the distribution of the maximum reaction force for the SENS benchmark with a single analysis step at time step $A$. The spread ($2\sigma$) reduced while the reference solution still lies in the range of the the posterior covariance.}
    \label{fig:2dmaxKDE}
  \end{minipage}
\end{figure}

An important quantity for the evaluation of the effect of a certain crack path is the remaining strength of a structure. A different path can potentially lead to a different residual load bearing capacity, which could in practice be an important quantity to base decisions on. Naturally, each random sample in the EnKF ensemble shows a different reaction force per time step. Figure \ref{fig:2dreact} shows the whole time line for the 2D SENS example. 
Regardless of the given initial damage, all samples behave similarly in the first, elastic only, part of the pseudo time line. Only when the fracture driving energy is large enough to initiate a crack that reaches the initial damage, differences become apparent. The spread of different reaction forces is visible in the plotted samples. At time step A, the EnKF analysis step is performed. A couple of interesting phenomena can be observed at this point: 
\begin{itemize}
    \item Firstly, the spread reduces.  While there are some outliers and a noticeable but smaller spread, most ensemble members now show a similar reaction force.
    \item Further in time, the average reaction force still closely follows the reference, i.e. the solution from which sparse data are drawn. Notice that the spread increases over time also for the updated ensemble. For time step E, Figure \ref{fig:2dreactKDE} shows a histogram of the updated ensemble compared to the same ensemble simulation without the Kalman analysis step. The standard deviation of the resulting distribution is smaller than the one for the case with no analysis step at A. Also, the reference value is still captured within the PDF.
    \item Figure \ref{fig:2dmaxKDE} captures the histogram of the peak force of each sample, i.e. the maximum load that the structure can resist. Also here, the variance reduces significantly compared to the prior.
\end{itemize}

\begin{figure}[ht!]
    \centering
    \includegraphics[width=10cm]{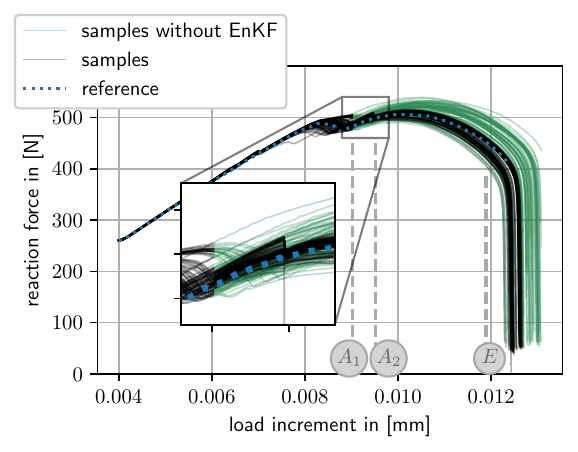}
    \caption{Reaction forces for the SENS benchmark with with two analysis steps at timesteps $A_1$ and $A_2$, visible in black. After the Kalman shift, the ensemble members cluster more closely around the reference solution. For comparison, the ensemble without Kalman shift is visible in green. } 
    
    \label{fig:2dreact2up}
\end{figure}

\begin{figure}[ht!]
    \begin{minipage}[b]{.45\linewidth}
    \centering
     \includegraphics[]{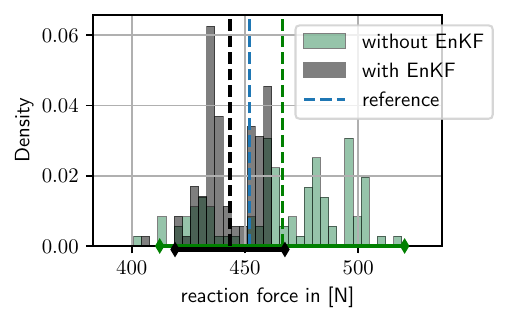}
    \caption{Histogram of the distribution of reaction forces at extrapolation time step $E$ for the SENS benchmark with two analysis steps at timesteps $A_1$ and $A_2$. The spread ($2\sigma$) reduced while the reference solution still lies in the range of the the posterior covariance.}
    \label{fig:2dreactKDE2upearly}
  \end{minipage}\hfill
  \begin{minipage}[b]{.45\linewidth}
    \centering
\includegraphics[]{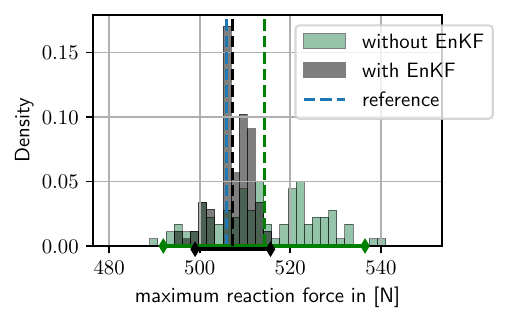}
    \caption{Histogram of the distribution of the maximum reaction force for the SENS benchmark with two analysis steps at timesteps $A_1$ and $A_2$. The spread ($2\sigma$) reduced while the reference solution still lies in the range of the the posterior covariance.}
    \label{fig:2dmaxKDE2upearly}
  \end{minipage}
\end{figure}

In Figure \ref{fig:2dreact2up}, two analysis steps are introduced, in alignment with Figure \ref{fig:stateUpdates}. 
Notice that, compared to only performing the Kalman shift once, the variance decreases even further. This is also visible in the histograms of the extrapolation time step $E$ and the histogram of the peak force in Figures \ref{fig:2dreactKDE2upearly} and \ref{fig:2dmaxKDE2upearly}, respectively.

%


\section{Discussion}
\label{sec:discussion}

Recalling the objective function of the ensemble Kalman update
\begin{equation}
        J_n(\bm{a}^\text{A}_{n,i}) = \frac{1}{2} \left| \bm{y}_{n} - \bm{H} \bm{a}^\text{A}_{n,i} \right|_{\Gamma}^2 + \frac{1}{2} \left| \bm{a}^\text{A}_{n,i} - \bm{a}^\text{F}_{n,i} \right|_{\bm C_{n}}^2,
\end{equation}
reveals that the physical model only enters mildly through the simulation of the prior ensemble. Indeed, the ensemble Kalman update is not forced to respect the physical constraints of the phase field model, which is the origin of the stability problems observed in the numerical experiments. The regularization projects the data-driven ensemble update back towards the physically admissible manifold, while also filtering spurious oscillations through an increased length-scale. In this section we will interpret the regularization in terms of proximal correction and formulate an outlook towards strongly constrained variational data assimilation in this context. 

To this end, it is instructive to adopt the perspective of energy minimization on the discrete level, see Section \ref{sec2}.
From this viewpoint, we can recast the regularization step of Section~\ref{sec:reg} in the following way: given an EnKF-assimilated state $(\boldsymbol{u}^{\text{A}}_{n,i},\boldsymbol{d}^{\text{A}}_{n,i})$, consider
\begin{equation}
    \tilde{\boldsymbol{d}}_{n,i}^\text{R} = \text{argmin}_{\boldsymbol{d}}\left(\tau \|  \boldsymbol{d} - \boldsymbol{d}^{\text{A}}_{n,i} \|_{\boldsymbol{C}_{\boldsymbol{d}}^{}} +  E_{h;n} (\boldsymbol{u}^{\text{A}}_{n,i},\boldsymbol{d};\boldsymbol{0},L)\right),
\end{equation}
where the irreversibility of cracks is lifted by setting $\boldsymbol{\varphi}' = \boldsymbol{0}$ and the phase fields are smoothed by employing $L > \ell$. This proximal step projects the purely data-driven Kalman shift back into the direction of the physical model and the parameter $\tau$ controls the balance between data and physics contribution. Moreover, the covariance matrix of the ensemble Kalman update naturally gives high weight to the ensemble parts with low variance. Nevertheless, as storing a complete covariance matrix in memory requires a lot of memory, we replace it with the identity matrix in the examples presented in this paper.


Having highlighted the prominent role of the physical model in phase field data assimilation it appears natural to target strongly constrained variational data assimilation in future work. The 4DVAR approach, for instance, considers smoothing instead of filtering over a time window with $N_T+1$ steps with the model entering as strong constraint. The respective optimization problem seeks to determine the solution vector at the beginning of that window $\bm{a}^\text{A}_{0}$ such that
    \begin{equation}
        J_0(\bm{a}^\text{A}_{0}) = \frac{1}{2} \sum_{n=0}^{N_T-1} \left| \boldsymbol{y}_{n+1} - \boldsymbol{H}  \tilde{\bm{a}}^\text{A}_{n+1} \right|_{\boldsymbol{\Gamma}}^2 +  \frac{1}{2} \left| \bm{a}^\text{A}_{0} - \boldsymbol{m}_0 \right|_{\boldsymbol{C}_0}^2,
    \end{equation}
    subject to the phase field update
    \begin{equation}
        \tilde{\bm{a}}^\text{A}_{n+1} = \mathcal{M}(\tilde{\bm{a}}^\text{A}_{n}),
    \end{equation}
    is minimized. Here, $\boldsymbol{m}_0$ is an initial guess, e.g., the forecast $\bm{a}^\text{F}_{n}$ at that time step. Part of the optimization problem are all time steps with data inside the assimilation window. This implies that in each iteration of the optimizer, the model has to be solved from time $0$ up until time $N_T$, so that the objective function can be evaluated. 
    Although 4DVAR is formulated for a mean vector instead of an ensemble, one could also replace the mean at time $0$ with an ensemble member and the covariance $\boldsymbol{C}_0$ with the ensemble approximated covariance.
    
    In principle, the assimilation window could be reduced to a single time step, thereby enforcing the model to hold for the analysis vector and thus only permitting physically admissible solutions. A drawback is that no closed form solution exists for the 4DVAR optimization problem and hence, gradient descent methods or others need to be employed. This in turn leads to a huge computational cost because the model needs to be solved multiple times per time step in the assimilation window and also gradients are typically required.


\section{Conclusion and Outlook}
In this paper, we presented a way to quantify and reduce uncertainty in a phase field brittle fracture problem by means of data assimilation. Random initial conditions lead to randomness in the crack path and hence also the remaining stiffness of the structure under consideration. In practice, knowledge over both these quantities is desired. Therefore, reducing the uncertainty about it by using an Ensemble Kalman Filter, given sparse measurement data, makes it possible to make more precise statements about the load capacity of a given structure compared to a stochastic model or data alone. 

The stochastic phase field model is propagated in time up until a time step at which data are available. At that point, the measurement data are assimilated with the model and the result is regularized in order to form a suitable initial condition for the next pseudo time step. The procedure was analysed in detail along with an illustrative 1D example. A 2D numerical benchmark example served to demonstrate the applicability of the method in more relevant settings. It could be demonstrated that the displacement and phase field states can be updated very accurately. 
However, a plain Ensemble Kalman Filter doesn't lead to acceptable results: A subsequent regularization is necessary to provide physically meaningful solutions.
The problem of a lack of regularization can potentially be solved by adopting an variational perspective on data assimilation and using methods such as 4DVar, which  constrain the analysis to a feasible model output without explicitly projecting the analysis after inference. However, an approach like that requires solving the model again for multiple time steps during inference, which leads to much higher computational cost.

Also, different sources of uncertainty such as randomness in parameters and the mesh should be addressed, as well as an extension of the method toward dynamical phase field methods. These involve derivatives in time for which the EnKF needs to be adapted.


\appendix

\section{Covariance Inflation and Localization}
\label{app1}
An EnKF is usually not used in the large ensemble limit, i.e. with more ensemble members than degrees of freedom. In the present examples, much fewer ensemble members are used compared to the high number of degrees of freedom in the phase field problem, which are necessary to resolve the crack. This can lead to numerical issues such as spurious long-distance correlations and filter collapse due to the rank deficiency of the empirical covariance matrix \cite{schillingsAnalysisEnsembleKalman2017}. Two countermeasures are commonly advised in the EnKF literature: Firstly, covariance inflation is used to ensure that the covariance doesn't vanish. As put forth in \cite{hamillAccountingErrorDue2005}, to inflate the covariance, each ensemble member is inflated individually before the analysis step. An inflation factor $r$, usually slightly greater than $1.0$, is used. For an ensemble member $[i]$ at time step $n$, we obtain
    \begin{equation}
        \bm{a}_{n,i}^{\text{inf}} = r ( \bm{a}_{n,i}  -   \bar{\bm{a}}_{n} ) +  \bar{\bm{a}}_{n}.
    \end{equation}
    From the inflated ensemble members, an inflated covariance matrix $\bm{C}_{n}^{\text{inf}}$ is computed.

    The second countermeasure is covariance localization which serves to prevent spurious correlations and hence, unphysical effects in the posterior state.
    According to e.g. \cite{houtekamerSequentialEnsembleKalman2001}, covariance localization is accomplished by performing the Schur product of a locally supported correlation matrix $\bm{\rho}^{\text{loc}}$ with the covariance matrix, i.e. 
    \begin{equation}
            \bm{C}_{n}^{\text{loc}}  = \bm{\rho}^{\text{loc}} \circ \bm{C}_{n}^{\text{loc}}.
    \end{equation}
    The authors in \cite{houtekamerSequentialEnsembleKalman2001} state that there are many suitable $\bm{\rho}^{\text{loc}}$, while a commonly used one has been defined by \cite{gaspariConstructionCorrelationFunctions1999} and closely resembles a squared exponential covariance. Hence, in this paper, we use a squared exponential covariance function for the localization which works well for the discussed problems.

\section*{Acknowledgments}
The source code which was developed along this paper will be made available on GitHub and Zenodo after acceptance of the paper. It is written mainly in Julia \cite{bezansonJuliaFreshApproach2017} with Ferrite.jl \cite{carlssonFerritejl2026} as a FEM backend.

\section*{Declarations}
\begin{itemize}
\item Funding\\
The research leading to these results received funding from the Deutsche Forschungsgemeinschaft (DFG, German Research Foundation) - Projektnummer 255042459 GRK2075/2.
\item Competing interests\\ 
The authors have no financial or proprietary interests in any material discussed in this article.
\item Data availability:\\
All data generated or analysed during this study are included in this published article.
\item Author contribution:\\
Conceptualization and Methodology: LH, UR, KAM, RJ; Formal analysis, investigation, validation and visualization: LH; Software: LH, KAM; Writing - original draft preparation: LH; Writing - review and editing: LH, UR, KAM, RJ; Funding acquisition, project administration and resources: UR, RJ; Supervision: UR, KAM, RJ.
\end{itemize}








\bibliographystyle{elsarticle-num}
\bibliography{ppkalman}

\end{document}